\documentclass[
reprint,
superscriptaddress,
amsmath,amssymb,
aps,
prb,
twocolumn
]{revtex4-2}

\usepackage{graphicx}
\usepackage{bm}
\usepackage[breaklinks]{hyperref}
\hypersetup{colorlinks=true, linkcolor=blue, citecolor=blue, filecolor=blue, urlcolor=blue}
\graphicspath{{figures}}
\usepackage{physics}
\usepackage{float}
\usepackage{dsfont}
\usepackage{enumitem}
\usepackage{xcolor}

\def\CH{\textcolor{magenta}}

\begin{document}

\title{Realization of decoherence-induced averaged symmetry-protected topological phases on quantum processors}

\author{Ruizhe Shen}
\email{e0554228@u.nus.edu}
\affiliation{Department of Physics, National University of Singapore, Singapore 117551}

\author{Xue-Jia Yu}
\email{xuejiayu@eitech.edu.cn}
\affiliation{Eastern Institute of Technology, Ningbo 315200, China}
\author{Ching Hua Lee}
\email{phylch@nus.edu.sg}
\affiliation{Department of Physics, National University of Singapore, Singapore 117551}
\date{\today}

\begin{abstract}
Symmetry-protected topological (SPT) phases are conventionally formulated for
pure states protected by exact symmetries. In open quantum systems, however,
decoherence generates mixed-state ensembles in which average symmetry can instead
emerge only after averaging over microscopic trajectories. In this work, we realize
decoherence-induced averaged SPT (ASPT) order on programmable quantum
processors. Starting from a one-dimensional cluster SPT, we engineer
sublattice-selective Pauli-$Z$ dephasing through ancilla-assisted quantum
circuits. We observe the
resulting symmetry conversion through the decay of the symmetry charge. Moreover, a two-replica
R\'enyi-2 string correlator measured through destructive SWAP readout remains
nontrivial, revealing the ASPT structure encoded at the density-matrix level. We further show that the engineered dephasing redistributes
spectral weight among reduced-stabilizer sectors while preserving the
characteristic twofold pairing of the half-chain entanglement spectrum. These results
establish a gate-based route to engineering and probing ASPT order and
demonstrate structured decoherence as a programmable resource for realizing
mixed-state topological quantum matter.
\end{abstract}
\maketitle

\section{Introduction}
\label{sec:introduction}
Symmetry-protected topological (SPT) phases provide an important principle for
quantum matter beyond the conventional Landau-Ginzberg-Wilson symmetry breaking paradigm
\cite{chenCompleteClassificationOneDimensional2011,
pollmannEntanglementSpectrumTopological2010,
pollmannSymmetryProtectionTopological2012,
chenSymmetryProtectedTopological2013,YU20261}.
An SPT phase is short-range entangled in the bulk, but cannot
be continuously deformed into a trivial product state without either closing the
gap or breaking the protecting symmetry
\cite{chenCompleteClassificationOneDimensional2011,
pollmannSymmetryProtectionTopological2012,
chenSymmetryProtectedTopological2013}.
Its nontrivial topology is reflected in several equivalent signatures, including
projective edge representations, characteristic degeneracies in the entanglement
spectrum, and nonlocal string order
\cite{pollmannEntanglementSpectrumTopological2010,
pollmannDetectionSymmetryProtected2012,
pollmannSymmetryProtectionTopological2012,
liHaldanePhaseS2008,kennedyHiddenSymmetryBreaking1992}.
These properties make SPT phases a useful setting for studying how symmetry,
entanglement, and topology can be detected in experimentally accessible quantum
many-body systems.

The experimental study of SPT phases has advanced rapidly with the development
of programmable quantum simulators and quantum processors
\cite{deLesleucExperimentalRealization2019,
tanRealizingSymmetryProtected2021,
zhangDigitalQuantumSimulation2022,
dumitrescuDynamicalTopologicalPhase2022,
shenRobustSimulationsManyBody2025,
penningtonPreparing100Qubit2026}, which have been effective in realizing a variety of non-unitary processes through post-selection or programmed measurements\cite{google2023measurement,chen2023high,smith2023deterministic,koh2023measurement,koh2026interacting,xu2026fractional}. Across superconducting processors,
trapped-ion systems, Rydberg arrays, and photonic platforms, recent experiments
have demonstrated the preparation and characterization of SPT states through
edge signatures, nonlocal string correlations, and dynamical topological
responses
\cite{deLesleucExperimentalRealization2019,
zhangDigitalQuantumSimulation2022,
dumitrescuDynamicalTopologicalPhase2022,
iqbalNonAbelianTopological2024,
tantivasadakarnLongRangeEntanglement2024,
sadouneLearningSymmetryProtected2024,
penningtonPreparing100Qubit2026}. Most existing SPT experiments, however,
remain formulated for pure states or approximately coherent dynamics
\cite{deLesleucExperimentalRealization2019,
zhangDigitalQuantumSimulation2022,
dumitrescuDynamicalTopologicalPhase2022,
shenRobustSimulationsManyBody2025,
penningtonPreparing100Qubit2026}, whereas noisy intermediate-scale quantum
devices and open quantum systems naturally generate mixed states through
decoherence, disorder, measurements, and classical averaging
\cite{nielsenQuantumComputationQuantum2010,
wildeQuantumInformationTheory2017,maAverageSymmetryProtected2023,
maTopologicalPhasesAverage2025}. In mixed states, symmetry has a richer
structure: a strong symmetry fixes the density matrix under left and right
action separately, while a weak, or average, symmetry requires only invariance
under conjugation by the symmetry operator
\cite{maAverageSymmetryProtected2023,maTopologicalPhasesAverage2025,
maSymmetryProtectedTopological2025,xueTensorNetworkFormulation2024,
guoLocallyPurifiedDensity2025,Guo2025PRXQuantum}. This distinction allows for
averaged SPT (ASPT) phases, in which the relevant symmetry need not be preserved
by each trajectory but is restored only after averaging over
disorder configurations or measurement branches
\cite{maAverageSymmetryProtected2023,maTopologicalPhasesAverage2025,
maSymmetryProtectedTopological2025,zhangQuantumCommunicationMixed2025,
leeSymmetryProtectedTopological2025,luNonequilibriumTopologicalResponse2026,
liGeneralizedSymmetryProtected2026}.

Very recently, an experimental realization of average-symmetry topology was
reported in a disordered Rydberg atom array, where structural disorder was used
to generate an average topological phase
\cite{yueAverageTopologicalPhase2026}. In that setting, the ensemble is formed
by different disorder realizations, and the topological signatures are extracted
from disorder-averaged observables
\cite{yueAverageTopologicalPhase2026}. This establishes the experimental
relevance of average-symmetry topology, but the ensemble itself is fixed by the
static disorder. A measurement-based realization is
conceptually distinct: rather than averaging over externally prepared disorder
configurations, the ensemble is generated intrinsically by quantum measurement
within programmable circuits. Each measurement outcome defines a
microscopic trajectory. This provides a programmable realization in which the microscopic trajectories
forming the ASPT ensemble are generated and resolved through measurement, while
the ensemble average is obtained by discarding the measurement record.

Here, we address this question by realizing a decoherence-induced averaged
cluster SPT on quantum processors
\cite{maAverageSymmetryProtected2023,maTopologicalPhasesAverage2025,
maSymmetryProtectedTopological2025,yueAverageTopologicalPhase2026}.
Starting from a one-dimensional cluster-state SPT
\cite{raussendorfOneWayQuantum2001,heinEntanglementGraphStates2004,
sonTopologicalOrderCluster2011,elseSymmetryProtectedPhases2012,
pollmannSymmetryProtectionTopological2012}, we engineer a local Pauli-$Z$
dephasing channel on the even sublattice using ancilla qubits
\cite{krausStatesEffectsOperations1983,
nielsenQuantumComputationQuantum2010}.
The ancilla measurement outcomes label the individual dephasing trajectories,
and averaging over these outcomes generates the mixed state used in the ASPT
protocol. This construction realizes a hybrid symmetry structure in which the
odd-sublattice symmetry remains strong, while the even-sublattice symmetry is
preserved only in the average sense at the level of the mixed-state density
matrix
\cite{maAverageSymmetryProtected2023,maTopologicalPhasesAverage2025,
maSymmetryProtectedTopological2025,xueTensorNetworkFormulation2024,
guoLocallyPurifiedDensity2025,zhangQuantumCommunicationMixed2025}.

We first measure the even-sublattice symmetry charge and observe its decay
under engineered dephasing, which characterizes the loss of a
strong-symmetry charge. We then show that the conventional single-copy string
order is suppressed by trajectory-dependent sign averaging and is therefore
insufficient to diagnose the ASPT state. To probe the ASPT structure at the
density-matrix level, we implement a two-replica doubled-string measurement
based on destructive SWAP readout, which yields the normalized R\'enyi-2 string
correlator
\cite{maSymmetryProtectedTopological2025}.
This diagnostic remains nontrivial even when the corresponding
single-copy string signal is strongly suppressed, revealing the mixed-state
ASPT structure associated with the average symmetry
\cite{maAverageSymmetryProtected2023,maTopologicalPhasesAverage2025,
xueTensorNetworkFormulation2024,guoLocallyPurifiedDensity2025}. We further reconstruct the half-chain reduced-density-matrix spectrum and show
that dephasing redistributes spectral weight among reduced-stabilizer sectors
while preserving the characteristic twofold pairing associated with the
projective edge degree of freedom
\cite{liHaldanePhaseS2008,pollmannEntanglementSpectrumTopological2010,
pollmannDetectionSymmetryProtected2012,pollmannSymmetryProtectionTopological2012}. Together, these measurements provide a broader characterization of how
engineered decoherence reshapes symmetry and topological matter on current programmable quantum hardware.

\begin{figure}
    \centering
    \includegraphics[width=0.99\linewidth]{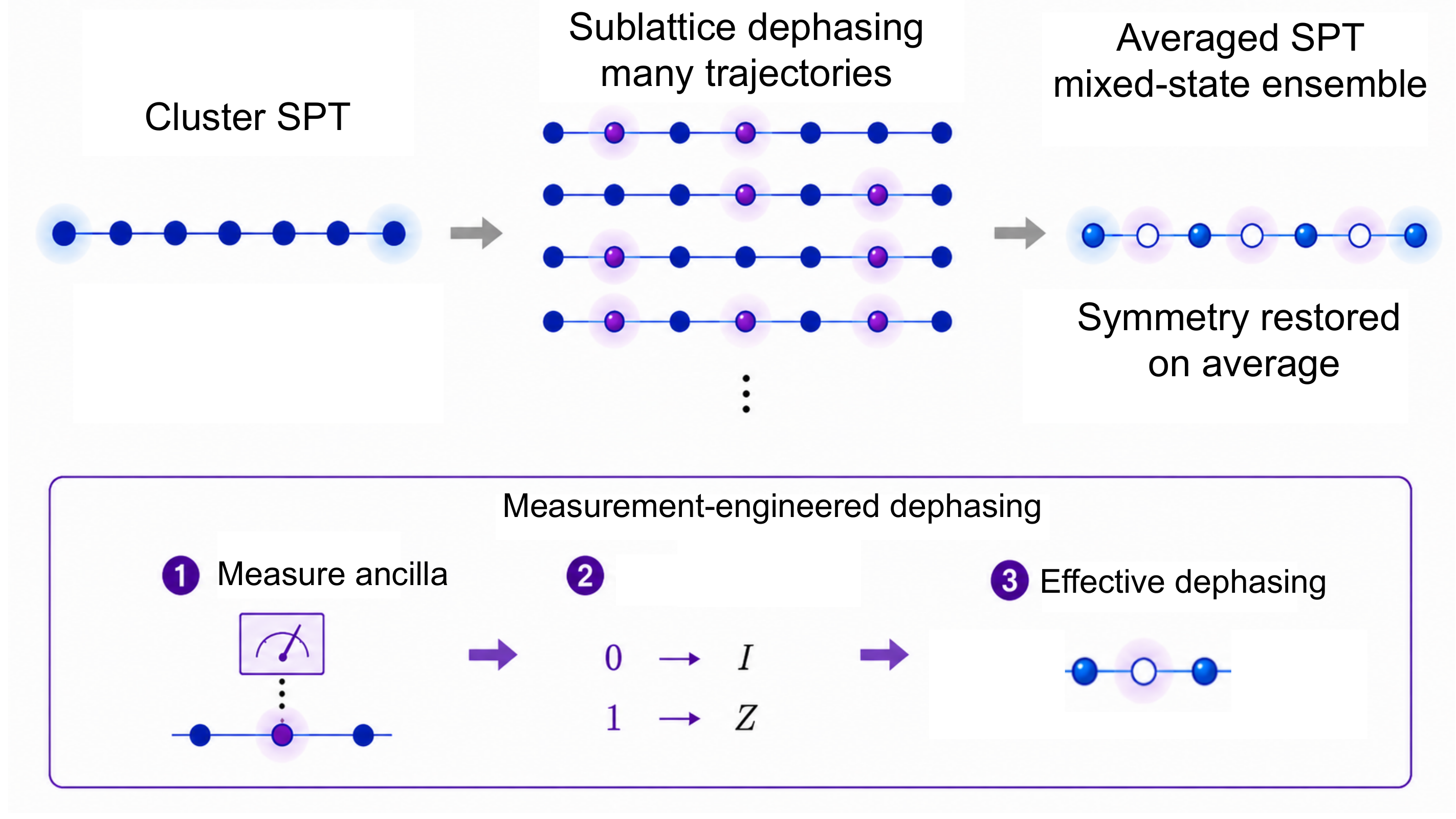}
    \caption{
Decoherence as a route to ASPT  order.
Starting from a pure cluster SPT with strong symmetry and protected edge/string order, local Pauli-$Z$ dephasing engineered by measurements is applied only to the even sublattice [see Eq.~\ref{eq:main_single_ancilla_branches}].
Each trajectory corresponds to a different probabilistic phase-flip pattern.
Averaging over these trajectories yields a mixed-state ensemble with weak, average symmetry and persistent topological order.
}
    \label{fig:fig1_SPT}
\end{figure}

\subsection{Averaged SPT states}

SPT phases are generally formulated for pure many-body states protected by a
symmetry group $G$
\cite{chenCompleteClassificationOneDimensional2011,
pollmannEntanglementSpectrumTopological2010,
pollmannSymmetryProtectionTopological2012,
chenSymmetryProtectedTopological2013}.
In one dimension, a pure SPT phase is a short-range-entangled phase that cannot
be connected to a trivial product state by a symmetry-preserving local unitary
\cite{chenCompleteClassificationOneDimensional2011,
pollmannSymmetryProtectionTopological2012,
chenSymmetryProtectedTopological2013}.
For a pure state $\rho_\psi=|\psi\rangle\langle\psi|$ in a trivial symmetry
sector, symmetry invariance imposes the strong condition
\begin{equation}
U_g\rho_\psi
=
\rho_\psi U_g
=
\rho_\psi,
\qquad
g\in G,
\label{eq:main_pure_strong_symmetry}
\end{equation}
where $U_g$ is the global symmetry operator. 

For mixed states, however, symmetry can be realized more generally at the
ensemble level. Such an ASPT state can be described through
\begin{equation}
\rho
=
\sum_\eta P(\eta)
|\psi_\eta\rangle\langle\psi_\eta|,
\label{eq:main_aspt_ensemble}
\end{equation}
where $\eta$ labels a trajectory, disorder configuration, decoherence history,
or measurement branch
\cite{maAverageSymmetryProtected2023,maTopologicalPhasesAverage2025,
leeSymmetryProtectedTopological2025,zhangQuantumCommunicationMixed2025,
shahInstabilitySteadyState2025}.
The symmetry need not fix each branch in the strong sense of
Eq.~\eqref{eq:main_pure_strong_symmetry}. Instead, it may act nontrivially on
the ensemble,
\begin{equation}
U_g|\psi_\eta\rangle
=
e^{i\varphi_g(\eta)}
|\psi_{g\eta}\rangle,
\label{eq:main_branch_symmetry_action}
\end{equation}
where $\varphi_g(\eta)$ is a branch-dependent phase. If the ensemble
probability is invariant under this action, the averaged
density matrix satisfies $U_g\rho U_g^\dagger
=
\rho$. 
This defines a weak, or averaged symmetry: the symmetry need not be respected
by each microscopic realization, but is restored after ensemble averaging
\cite{maAverageSymmetryProtected2023,maTopologicalPhasesAverage2025,
maSymmetryProtectedTopological2025,xueTensorNetworkFormulation2024,
guoLocallyPurifiedDensity2025}.

In this work, we focus on a special class of SPT states known as cluster states.
The one-dimensional cluster state $|\psi_{\rm cl}\rangle$ is defined by the
stabilizers
\begin{equation}
K_j=Z_{j-1}X_jZ_{j+1},
\qquad
K_j|\psi_{\rm cl}\rangle=|\psi_{\rm cl}\rangle,
\label{eq:main_cluster_stabilizer}
\end{equation}
with $\rho_{\rm cl}=|\psi_{\rm cl}\rangle\langle\psi_{\rm cl}|$ [see Eq.~\ref{eq:main_cluster_preparation} below].  Here, $X_j$ and $Z_j$ denote the
single-qubit Pauli operators acting on site $j$. Such a state realizes
an SPT phase protected by the
$\mathbb{Z}_2^{\rm odd}\times\mathbb{Z}_2^{\rm even}$ symmetry, generated by
\begin{equation}
U_{\rm odd}=\prod_{j\in{\rm odd}}X_j,
\qquad
U_{\rm even}=\prod_{j\in{\rm even}}X_j.
\label{eq:main_cluster_symmetry}
\end{equation}
This pure cluster state is an exactly solvable representative, or fixed-point
state, meaning that it captures the
characteristic SPT structure in its simplest form. For endpoints $i$ and $k$
on the same sublattice, the corresponding cluster string operator is
\begin{equation}
\mathcal{O}_{i,k}
=
Z_i
\left(
\prod_{r=i+1,i+3,\ldots,k-1}
X_r
\right)
Z_k ,
\label{eq:main_cluster_string}
\end{equation}
which satisfies $\langle\mathcal{O}_{i,k}\rangle=1$ in the ideal
cluster state
\cite{kennedyHiddenSymmetryBreaking1992,
pollmannEntanglementSpectrumTopological2010,pollmannDetectionSymmetryProtected2012,
sonTopologicalOrderCluster2011,elseSymmetryProtectedPhases2012}.

For the cluster state in the $+1$ symmetry sector, both sublattice symmetries
are strong,
\begin{equation}
U_\alpha\rho_{\rm cl}
=
\rho_{\rm cl}U_\alpha
=
\rho_{\rm cl},
\qquad
\alpha\in\{{\rm odd},{\rm even}\}.
\label{eq:main_cluster_strong_symmetry}
\end{equation}
The cluster state therefore provides a pure-state reference point from which an
ASPT state can be generated by decoherence, disorder averaging, or
measurement-conditioned ensembles
\cite{maAverageSymmetryProtected2023,maTopologicalPhasesAverage2025,
maSymmetryProtectedTopological2025,xueTensorNetworkFormulation2024,
guoLocallyPurifiedDensity2025,leeSymmetryProtectedTopological2025,
zhangQuantumCommunicationMixed2025,shahInstabilitySteadyState2025,
guoStrongWeakSpontaneous2024,kunoStrongWeakSpontaneous2025,
liGeneralizedSymmetryProtected2026,liMixedStateTopological2026}.
Mixed-state SPT phases are more general and need not be ASPT phases (see Appendix.~\ref{sec:methods_feedback_mspt} for more details). Here, the ASPT  structure arises because one component of the original
$\mathbb{Z}_2^{\rm odd}\times\mathbb{Z}_2^{\rm even}$ symmetry remains
strong, while the other survives only as a weak symmetry after ensemble
averaging.

\subsection{Decoherence-induced averaged cluster state}
Our construction of the averaged cluster SPT state is based on converting one
component of the protecting
$\mathbb{Z}_2^{\rm odd}\times\mathbb{Z}_2^{\rm even}$ symmetry from a strong
symmetry to an average symmetry through engineered decoherence. Specifically,
we apply local Pauli-$Z$ dephasing only to the even sublattice
\cite{nielsenQuantumComputationQuantum2010,
wildeQuantumInformationTheory2017,maAverageSymmetryProtected2023,
maTopologicalPhasesAverage2025,maSymmetryProtectedTopological2025,
zhangQuantumCommunicationMixed2025,luNonequilibriumTopologicalResponse2026}.
 The local channel is
\begin{equation}
\mathcal{E}_j^Z(\rho)
=
(1-p)\rho
+
pZ_j\rho Z_j,
\qquad
j\in {\rm even},
\label{eq:main_local_dephasing}
\end{equation}
where $p$ is the probability of the local $Z_j$ branch.  Applying this channel
independently to all even sites gives
\begin{equation}
\rho_{\rm dec}
=
\prod_{j\in {\rm even}}
\mathcal{E}_j^Z(\rho_{\rm cl}) .
\label{eq:main_decohered_state}
\end{equation}
This is the channel-level route from the pure cluster-state SPT fixed point to a
mixed-state ensemble with weak, or average, symmetry
\cite{maAverageSymmetryProtected2023,maTopologicalPhasesAverage2025,
maSymmetryProtectedTopological2025,xueTensorNetworkFormulation2024,
guoLocallyPurifiedDensity2025,leeSymmetryProtectedTopological2025,
zhangQuantumCommunicationMixed2025,shahInstabilitySteadyState2025,
guoStrongWeakSpontaneous2024,kunoStrongWeakSpontaneous2025,
liGeneralizedSymmetryProtected2026,liMixedStateTopological2026}.

Under the above dephasing channels, the odd-sublattice symmetry remains strong,
whereas the even-sublattice symmetry survives only at the ensemble level: $U_{\rm odd}\rho_{\rm dec}
=
\rho_{\rm dec}U_{\rm odd}
=
\rho_{\rm dec},
U_{\rm even}\rho_{\rm dec}U_{\rm even}^\dagger
=
\rho_{\rm dec},$ (see Appendix.~\ref{sec:supp_z_dephasing_aspt_justification} for more details)
The engineered mixed state of Eq.~\eqref{eq:main_decohered_state} therefore
realizes the hybrid strong--average symmetry structure
[Fig.~\ref{fig:fig1_SPT}]
\begin{equation}
\mathbb{Z}_2^{{\rm odd},{\rm strong}}
\times
\mathbb{Z}_2^{{\rm even},{\rm average}}.
\label{eq:main_strong_weak_structure}
\end{equation}
Thus, the even-sublattice symmetry is not destroyed by dephasing, but changes
from a strong symmetry of the pure cluster state to an average symmetry of the
mixed-state ensemble (see Appendix.~\ref{sec:supp_z_dephasing_aspt_justification} for more details). This strong-to-average conversion underlies the
decoherence-induced averaged cluster SPT state realized in this work.

\begin{figure*}
    \centering
    \includegraphics[width=0.9\linewidth]{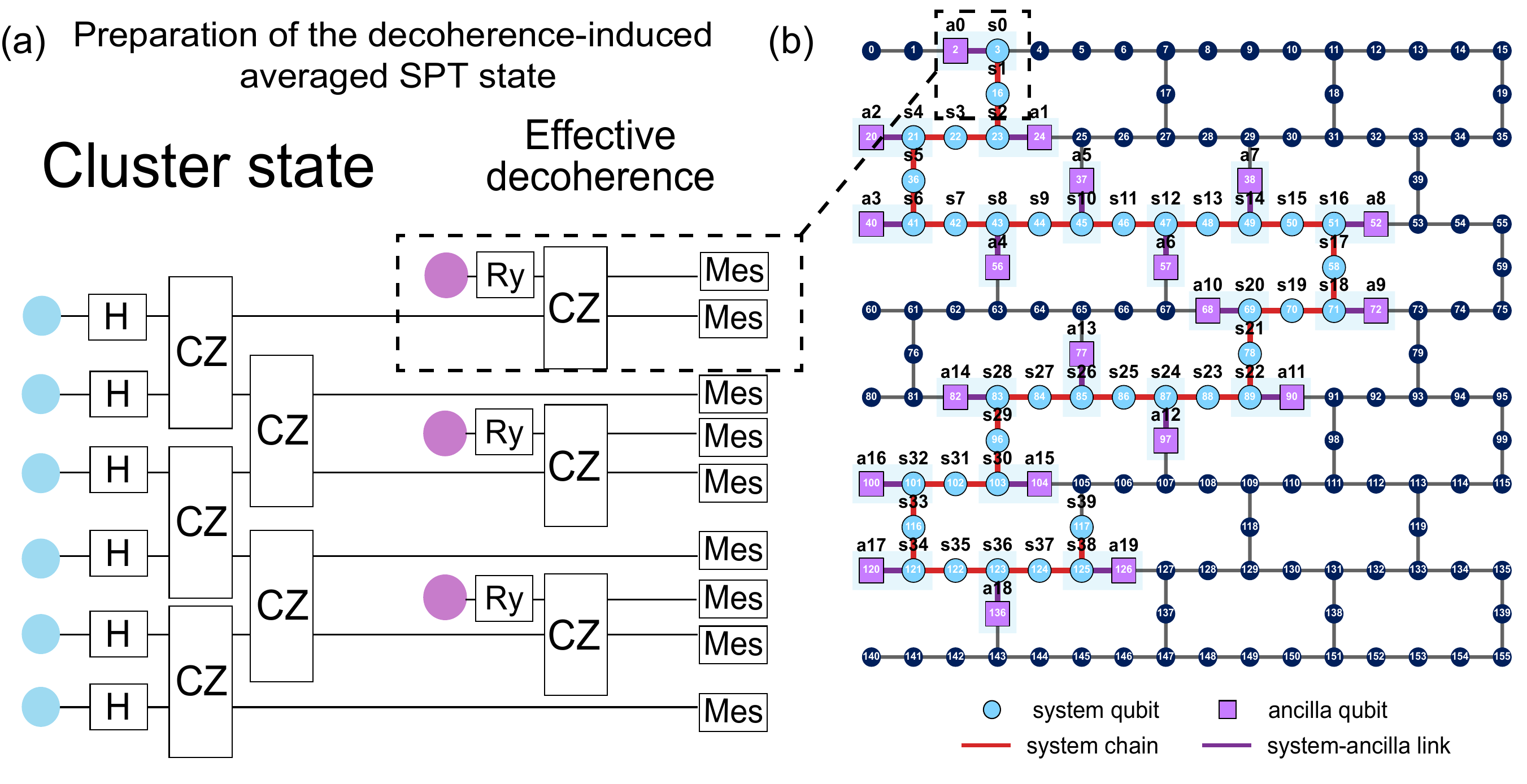}
\caption{
\textbf{Circuit architecture for realizing ASPT order on IBM Quantum hardware.}
(a), Preparation of the decoherence-induced ASPT state [Eq.~\ref{eq:main_protocol_summary}]. The one-dimensional cluster state is first prepared using Hadamard and nearest-neighbor CZ gates [Eq.~\ref{eq:main_cluster_preparation}], followed by local system--ancilla couplings that realize effective Pauli-$Z$ dephasing on the even sublattice [see Fig.~\ref{fig:fig1_SPT}]. Measuring the ancillas records the corresponding dephasing trajectory [Eq.~\ref{eq:main_local_dephasing}].
(b), Hardware embedding of the circuit in (a). Cyan and purple sites denote system and ancilla qubits, respectively; red links indicate the one-dimensional system chain, whereas purple links denote the system--ancilla couplings used to implement the engineered decoherence. 
}
    \label{fig:device}
\end{figure*}

\subsection{{Quantum Circuits for preparing ASPT  states}}\label{sec:main_circuit_realization}

At the circuit level, cluster states can be readily prepared using CZ entangling gates between neighboring qubits \cite{raussendorfOneWayQuantum2001,raussendorfMeasurementBasedQuantum2003,
heinEntanglementGraphStates2004,heinEntanglementGraphStatesApplications2006,
sonTopologicalOrderCluster2011,elseSymmetryProtectedPhases2012,
pollmannEntanglementSpectrumTopological2010,pollmannSymmetryProtectionTopological2012}. This setup is shown in Fig.~\ref{fig:device} (a):  For system qubits, starting from $\ket{0}^{\otimes L}$, a layer of Hadamard gates prepares
$\ket{+}^{\otimes L}$, and nearest-neighbour controlled-$Z$ gates generate
\begin{equation}
\ket{\psi_{\rm cl}}
=
\left(
\prod_{j=1}^{L-1}
{\rm CZ}_{j,j+1}
\right)
\ket{+}^{\otimes L}.
\label{eq:main_cluster_preparation}
\end{equation}
Because all controlled-$Z$ gates commute, the entangling layer can be scheduled
as two non-overlapping nearest-neighbour sublayers.

The decoherence-induced ASPT state in Eq.~\eqref{eq:main_decohered_state}
cannot be generated from the pure cluster state by a unitary acting on the
system alone. Instead, the desired
mixed state is obtained by coupling the system to ancilla degrees of freedom
and subsequently averaging over the ancilla measurement
outcomes
\cite{krausStatesEffectsOperations1983,nielsenQuantumComputationQuantum2010,
wildeQuantumInformationTheory2017}.  The corresponding circuit and hardware
layouts are shown in Figs.~\ref{fig:device}(a,b). The cyan qubits encode the
one-dimensional cluster chain, while ancilla qubits are coupled to the even
system sites to realize the local dephasing channels
\cite{raussendorfOneWayQuantum2001,raussendorfMeasurementBasedQuantum2003,
heinEntanglementGraphStates2004,sonTopologicalOrderCluster2011,
maAverageSymmetryProtected2023,maTopologicalPhasesAverage2025,
maSymmetryProtectedTopological2025,zhangQuantumCommunicationMixed2025, 
luNonequilibriumTopologicalResponse2026}.  
The experimental state preparation sequence is
\begin{equation}
\ket{0}^{\otimes L}
\longrightarrow
\ket{+}^{\otimes L}
\longrightarrow
\ket{\psi_{\rm cl}}
\longrightarrow
\rho_{\rm dec},
\label{eq:main_protocol_summary}
\end{equation}
where the last arrow denotes ancilla-generated dephasing followed by averaging
over measurement outcomes.

For each even system qubit $q_j$, an ancilla $a_j$ is initialized in
$|0\rangle$ and rotated according to
\begin{equation}
R_y(2\theta)|0\rangle_{a_j}
=
\sqrt{1-p}\,|0\rangle_{a_j}
+
\sqrt{p}\,|1\rangle_{a_j},
\qquad
p=\sin^2\theta .
\label{eq:main_ancilla_preparation}
\end{equation}
For an arbitrary system state $|\psi\rangle$, the system--ancilla state
after the CZ coupling is
\begin{equation}
|\Psi_j\rangle
=
\sqrt{1-p}\,
|\psi\rangle|0\rangle_{a_j}
+
\sqrt{p}\,
Z_j|\psi\rangle|1\rangle_{a_j}.
\label{eq:main_single_ancilla_dilation}
\end{equation}
A measurement of $a_j$ in the computational basis produces the two conditional
system states $|\psi\rangle$ and $Z_j|\psi\rangle$ with probabilities $1-p$
and $p$, respectively. The joint
system--ancilla state is therefore [see the schematic illustration shown in Fig.~\ref{fig:fig1_SPT}]
\begin{equation}
\rho_{Sa_j}^{\rm meas}
=
(1-p)\rho\otimes|0\rangle\langle0|_{a_j}
+
pZ_j\rho Z_j\otimes|1\rangle\langle1|_{a_j}.
\label{eq:main_single_ancilla_branches}
\end{equation}
Averaging over the ancilla outcome, equivalently tracing out the ancilla, gives
\begin{equation}
{\rm Tr}_{a_j}\!\left[\rho_{Sa_j}^{\rm meas}\right]
=
(1-p)\rho+pZ_j\rho Z_j
=
\mathcal{E}_j^Z(\rho),
\label{eq:main_ancilla_dephasing_derivation}
\end{equation}
which reproduces the local dephasing channel given in
Eq.~\eqref{eq:main_local_dephasing} (see Appendix.~\ref{sec:methods_ancilla_dephasing} for more details). Applying the construction independently to all even sites gives
\begin{equation}
\rho_{SA}^{\rm meas}
=
\sum_{\bm{\eta}}
P(\bm{\eta})
Z(\bm{\eta})\rho_{\rm cl}Z(\bm{\eta})
\otimes
|\bm{\eta}\rangle\langle\bm{\eta}|,
\label{eq:main_full_ancilla_dilation}
\end{equation}
where
$Z(\bm{\eta})=\prod_{j\in{\rm even}}Z_j^{\eta_j}$ and
$P(\bm{\eta})=\prod_{j\in{\rm even}}p^{\eta_j}(1-p)^{1-\eta_j}$.
Discarding the ancilla measurement record yields the mixed state $\rho_{\rm dec}
=
{\rm Tr}_{\rm a}\!\left[\rho_{SA}^{\rm meas}\right]
=
\sum_{\bm{\eta}}
P(\bm{\eta})
Z(\bm{\eta})\rho_{\rm cl}Z(\bm{\eta})$. 
A conventional single-copy observable therefore measures only the linear
average over these trajectories,
\begin{equation}
\langle O\rangle_{\rm dec}
=
{\rm Tr}(\rho_{\rm dec}O)
=
\sum_{\bm{\eta}}
P(\bm{\eta})\,\langle O\rangle_{\bm{\eta}},
\end{equation}
where
$\langle O\rangle_{\bm{\eta}}
=
{\rm Tr}[Z(\bm{\eta})\rho_{\rm cl}Z(\bm{\eta})O]$.
For Pauli-string observables, the dephasing trajectories can contribute with
opposite signs depending on $Z(\bm{\eta})$.
These opposite-sign contributions may cancel, and
consequently, the suppression of a single-copy observable can reflect
trajectory averaging rather than the disappearance of the underlying ASPT
order.

\begin{figure*}
    \centering
    \includegraphics[width=0.8\linewidth]{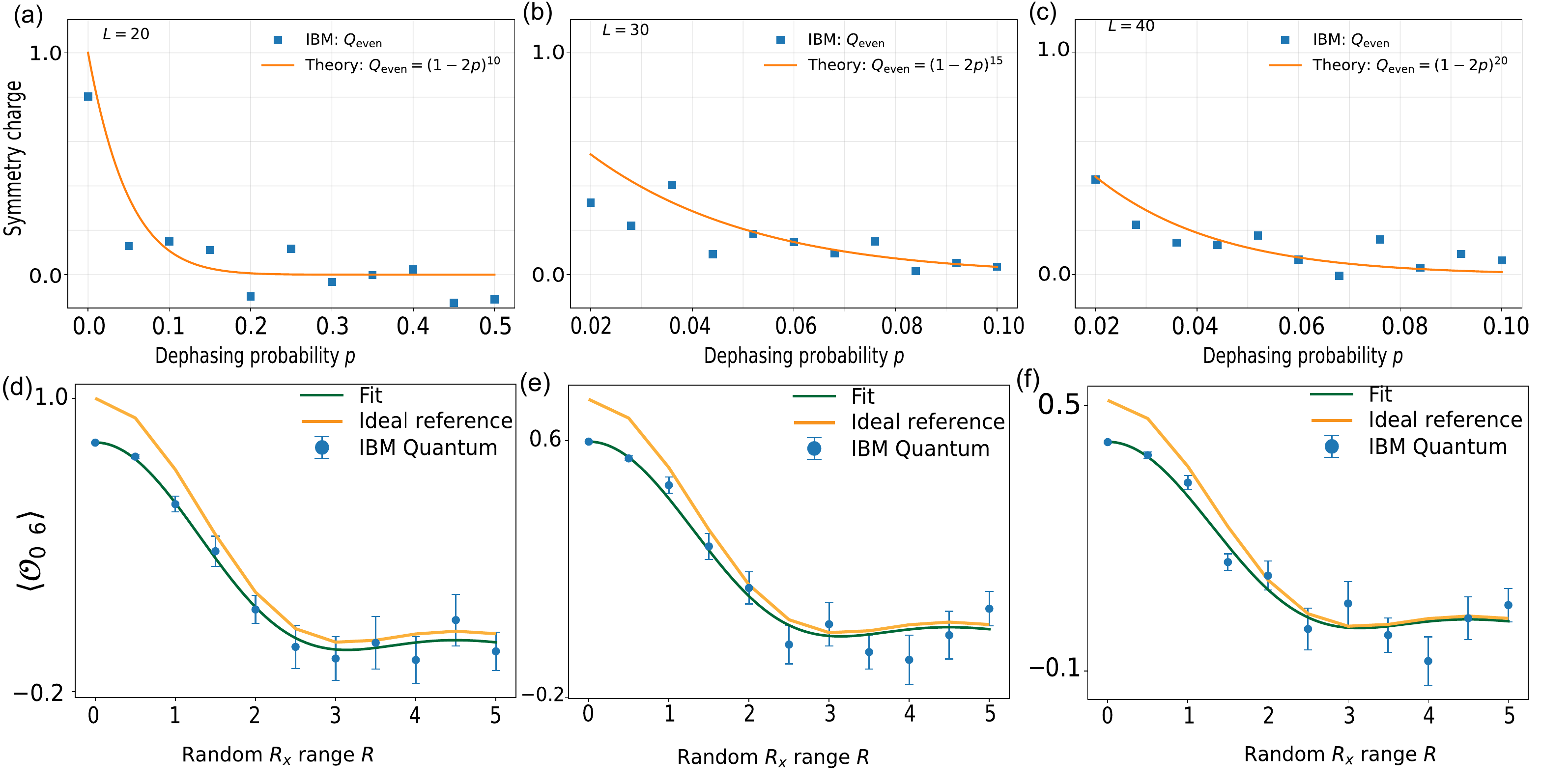}
\caption{{Quantum hardware measurements of decoherence-induced symmetry-charge and single-copy string-order decay. }
(a--c) Measurement of the even-sublattice symmetry charge $Q_{\rm even}$ defined in Eq.~\ref{eq:main_qeven_definition} under engineered even-sublattice dephasing.
The panels show $Q_{\rm even}$ as a function of the dephasing probability $p$ for physical chain lengths
(a) $L=20$, (b) $L=30$, and (c) $L=40$.
Blue markers denote IBM Quantum data, and orange curves show the ideal finite-size prediction in Eq.~\ref{eq:main_symmetry_charges}.
The suppression of $Q_{\rm even}$ with increasing $p$ and $L$ reflects the conversion of the even-sublattice symmetry from a strong symmetry to the weak, ensemble-level symmetry described in Eq.~\ref{eq:main_strong_weak_structure}.
(d--f) Single-copy string-order measurements in the presence of spatially random coherent $R_x$ rotations and engineered dephasing, following the circuit protocol in Eq.~\ref{eq:main_random_rx_protocol}.
The measured observable is the cluster string operator $\mathcal{O}_{i,k}$ defined in Eq.~\ref{eq:main_cluster_string}.
Here we use a chain of length $L=12$ with endpoints $i=0$ and $k=6$, and plot the random-field-averaged string signal [Eq.~\ref{eq:main_random_rx_protocol}] as a function of the random rotation range $R$.
Panels correspond to dephasing probabilities
(d) $p=0$, (e) $p=0.05$, and (f) $p=0.10$.
Blue markers show IBM Quantum data.
Orange curves show the ideal analytic reference in Eq.~\ref{eq:main_random_rx_reference_average}, and green curves show fits to the IBM Quantum data.
The systematic suppression of the single-copy string signal with increasing
$R$ and $p$, in agreement with the prediction, shows that the
observed decay is primarily induced by the intentionally introduced coherent
rotations and engineered dephasing rather than by hardware noise.
}
    \label{fig:res2}
\end{figure*}

\section{Quantum hardware measurement results}
In this section, we present quantum-hardware measurements that characterize
the prepared decoherence-induced ASPT states at two levels. A key
feature of ASPT order is that its nontrivial structure is encoded in the
mixed-state density matrix and therefore cannot, in general, be established
from conventional observables. Observable-level measurements, such as the single-copy string correlator, are instead
used to characterize the effect of the engineered decoherence. To directly probe the ASPT structure at the density-matrix level, we
employ two complementary approaches: an effective two-copy measurement of the
R\'enyi-2 string correlator, and direct density-matrix reconstruction.

\subsection{Benchmark of decoherence effects}  
\label{sec:main_symmetry_charge}
We first implement the circuit shown in Fig.~\ref{fig:device}(a) and perform a
global symmetry-charge benchmark to probe the loss of an
even-sublattice symmetry charge under the engineered dephasing channel
\cite{maAverageSymmetryProtected2023,maTopologicalPhasesAverage2025,
maSymmetryProtectedTopological2025,zhangQuantumCommunicationMixed2025,
luNonequilibriumTopologicalResponse2026}.
This measurement diagnoses the suppression of the strong-symmetry component,
while the surviving average symmetry is a property of the ensemble-averaged
density matrix.

Here, we measure the even-sublattice
symmetry charge
\begin{equation}
Q_{\rm even}
=
\Tr(\rho_{\rm dec}U_{\rm even}),\qquad
U_{\rm even}
=
\prod_{j\in{\rm even}}X_j .
\label{eq:main_qeven_definition}
\end{equation}
For the pure cluster state in the $+1$ symmetry-charge sector,
$Q_{\rm even}=1$. Under dephasing, $Q_{\rm even}$ decreases as the state loses
its definite even-sublattice symmetry charge. Such decay therefore provides a
direct diagnostic of the loss of strong even-sublattice symmetry. However, $Q_{\rm even}\to0$ does not imply that the even-sublattice
symmetry itself is completely lost: the weak symmetry can remain intact at the
density-matrix level through
$U_{\rm even}\rho_{\rm dec}U_{\rm even}^{\dagger}=\rho_{\rm dec}$.

On IBM Quantum hardware, $Q_{\rm even}$ is measured by reading out all
even-sublattice system qubits in the $X$ basis 
\cite{raussendorfOneWayQuantum2001,heinEntanglementGraphStates2004,
nielsenQuantumComputationQuantum2010}. Operationally, Hadamard gates are
applied before computational-basis measurement, so that the product of the
measured $Z$-basis eigenvalues estimates the many-body symmetry operator
$U_{\rm even}$. Under dephasing, the theoretically predicted (ideal) scaling of the $Q_\text{even}$ charge follows directly from the adjoint action of the dephasing
channel. For a dephased even site,
\begin{equation}
\left(\mathcal{E}^{Z}_{j}\right)^\dagger(X_j)
=
(1-p)X_j+pZ_jX_jZ_j
=
(1-2p)X_j .
\label{eq:main_channel_adjoint_x}
\end{equation}
Thus each even-site $X_j$ factor in $U_{\rm even}$ contributes a multiplicative
factor of $1-2p$.  For an even-length chain with $L/2$ dephased even-sublattice
sites,
\begin{equation}
\left(\mathcal{E}^{Z}_{\rm even}\right)^\dagger(U_{\rm even})
=
(1-2p)^{L/2}U_{\rm even}.
\label{eq:main_adjoint_even_charge}
\end{equation}
Since the initial cluster state lies in the $+1$ sector of $U_{\rm even}$, the
ideal symmetry charge is
\begin{equation}
Q_{\rm even}
=
(1-2p)^{L/2}.
\label{eq:main_symmetry_charges}
\end{equation}
By contrast, the odd-sublattice generator has no $X$ support on the dephased
even sublattice and therefore remains a strong symmetry in the ideal channel.

As shown in Fig.~\ref{fig:device}, we select a hardware layout on the IBM
Heron processor that accommodates up to $L=40$ system qubits together with
$20$ ancilla qubits. Using this layout, we measure $Q_{\rm even}$ for chain
lengths $L=20$, $30$, and $40$, as shown in
Figs.~\ref{fig:res2}(a--c). For all system sizes, the symmetry charge decreases
with increasing dephasing probability $p$. The ideal prediction in
Eq.~\ref{eq:main_symmetry_charges} captures the expected finite-size scaling:
at fixed $p$, the suppression of $Q_{\rm even}$ becomes stronger with
increasing $L$, because each of the $L/2$ independently dephased
even-sublattice sites contributes a multiplicative factor to the global
symmetry charge.  This size-dependent decay demonstrates the progressive loss
of an even-sublattice symmetry charge, and hence the suppression of
the corresponding strong symmetry. Importantly, the decay of
$Q_{\rm even}$ does not imply that the even-sublattice symmetry is destroyed:
the engineered channel preserves it in the average sense.

Following the symmetry-charge measurements, we use the conventional single-copy
cluster string correlator as a controlled benchmark of the engineered
decoherence \cite{pollmannEntanglementSpectrumTopological2010,
pollmannDetectionSymmetryProtected2012,pollmannSymmetryProtectionTopological2012,
sonTopologicalOrderCluster2011,elseSymmetryProtectedPhases2012}.
For the pure cluster state,
$\langle\mathcal{O}_{i,j}\rangle=1$, whereas even-sublattice $Z$ dephasing
suppresses the string according to
\begin{equation}
\langle\mathcal{O}_{i,j}\rangle_{\rm dec}
=
(1-2p)^{N_{\rm dep}(i,j)},
\label{eq:main_string_ij_decay}
\end{equation}
where $N_{\rm dep}(i,j)$ counts the dephased even-sublattice sites on which
$\mathcal{O}_{i,j}$ has $X$ support. This decay originates from the averaging
of trajectory-dependent string signs and therefore does not, by itself, imply
the disappearance of the ASPT order
\cite{maAverageSymmetryProtected2023,maTopologicalPhasesAverage2025,
maSymmetryProtectedTopological2025,xueTensorNetworkFormulation2024,
guoLocallyPurifiedDensity2025}. {To distinguish this controlled suppression from uncontrolled hardware
noise, we introduce a tunable
coherent perturbation whose symmetry action and effect on the string correlator
can be estimated analytically. We choose spatially random $R_x$ rotations because they
commute with the protecting
$\mathbb{Z}_2^{\rm odd}\times\mathbb{Z}_2^{\rm even}$ symmetry generators and
therefore perturb the cluster state without explicitly breaking the symmetry.} Specifically, after preparing the cluster state, we apply a
spatially random layer
$U_R(\bm{\phi})=\prod_j R_x(\phi_j)$, with independently sampled
$\phi_j\in[0,R]$, followed by the same ancilla-assisted dephasing channel:
\begin{equation}
\rho_{\rm dec}(\bm{\phi})
=
\left(
\prod_{j\in{\rm even}}\mathcal{E}_j^Z
\right)
\left[
U_R(\bm{\phi})\rho_{\rm cl}U_R^\dagger(\bm{\phi})
\right].
\label{eq:main_random_rx_protocol}
\end{equation}
Averaging over the random rotations gives
$\mathbb{E}_{\phi}[\cos\phi]=\sin R/R$. The $R_x$ layer therefore suppresses
the two endpoint $Z$ operators of the cluster string by
$(\sin R/R)^2$, while the interior $X$ string is affected only by the
engineered dephasing. The ideal circuit prediction is consequently
\begin{equation}
\overline{\langle\mathcal{O}_{i,j}\rangle}_{\rm ideal}
\approx
\left(\frac{\sin R}{R}\right)^2
(1-2p)^{N_{\rm dep}(i,j)},
\label{eq:main_random_rx_reference_average}
\end{equation}
which at $R=0$, reduces to
Eq.~\eqref{eq:main_string_ij_decay}.

Figs.~\ref{fig:res2}(d--f) show the single-copy string correlator for an
$L=12$ chain as a function of the random-rotation range $R$ at several
dephasing probabilities. The IBM Quantum results follow the ideal-circuit
predictions closely, with the string signal decreasing systematically as either
$R$ or $p$ is increased. This agreement shows that the observed suppression is
primarily driven by the
engineered dephasing, rather than by hardware noise. At the same time, the decay of the conventional string correlator illustrates
its limitation as an ASPT diagnostic: a vanishing single-copy signal does not
distinguish an averaged SPT state from a topologically trivial mixed state. To
access the mixed-state order that remains hidden after trajectory averaging, we
therefore turn to a  two-replica diagnostic involving independently
prepared copies of $\rho_{\rm dec}$ in the next section
\cite{maAverageSymmetryProtected2023,maTopologicalPhasesAverage2025,
maSymmetryProtectedTopological2025,xueTensorNetworkFormulation2024,
guoLocallyPurifiedDensity2025}.

\begin{figure*}
    \centering
    \includegraphics[width=0.9\linewidth]{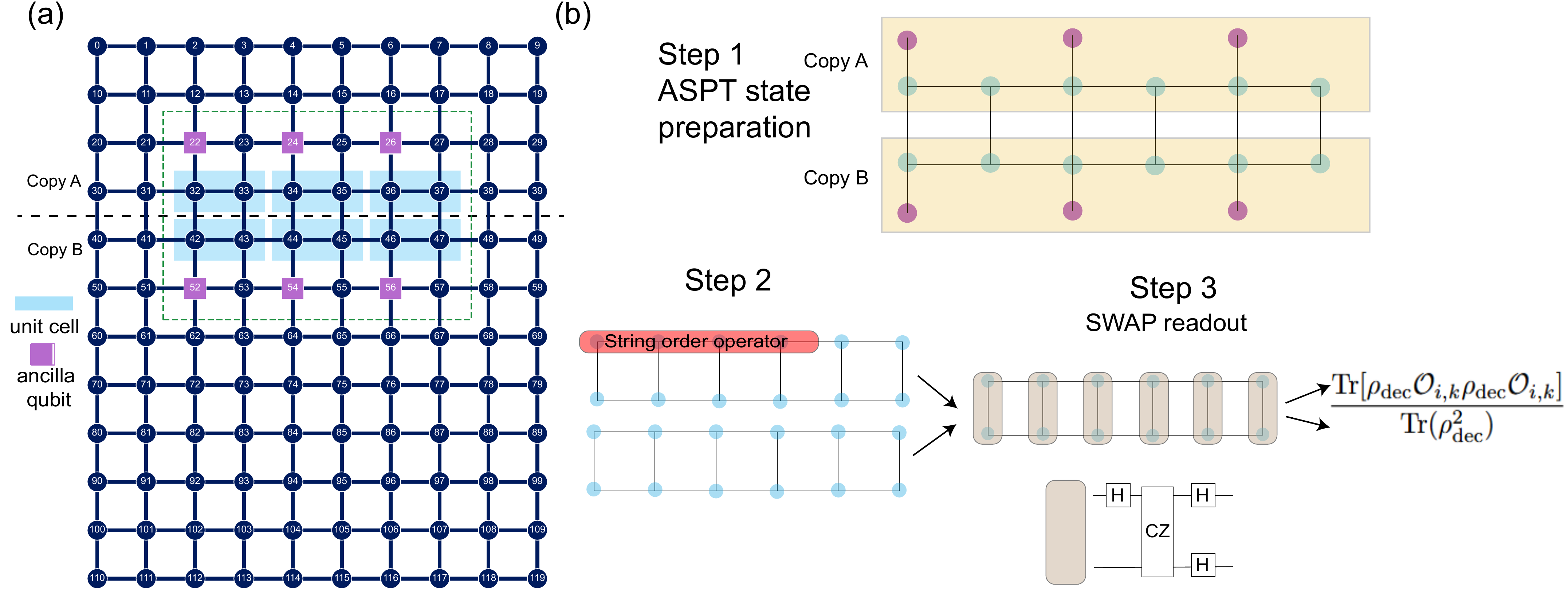}
\caption{
\textbf{Two-replica circuit for measuring the normalized R\'enyi-2 string diagnostic of the ASPT  phase.}
\textbf{a}, Hardware embedding of the two-replica experiment. Copies $A$ and $B$ are mapped onto two parallel system-qubit chains, with ancillary qubits coupled to the even sublattice to implement the engineered dephasing channel. The repeated unit-cell structure realizes the same ASPT preparation independently on the two replicas.
\textbf{b}, Three-step protocol for measuring the nonlinear string correlator $C_{i,k}^{(2)}$ defined in Eq.~\ref{eq:main_swap_ratio}. \textit{Step 1}: two independent copies of the decohered cluster state, $\rho_{\rm dec}^{A}$ and $\rho_{\rm dec}^{B}$, are prepared using identical  ASPT circuits, with independently sampled ancilla outcomes. \textit{Step 2}: for the numerator, the cluster string $\mathcal{O}_{i,k}$ (red shaded region) is applied to replica $A$ before the replica readout; omitting this insertion gives the denominator. These two measurements yield $N_{i,k}=\Tr(\rho_{\rm dec}\mathcal{O}_{i,k}\rho_{\rm dec}\mathcal{O}_{i,k})$ and $D=\Tr(\rho_{\rm dec}^{2})$, respectively [Eqs.~\ref{eq:main_swap_numerator} and \ref{eq:main_swap_denominator}]. \textit{Step 3}: a destructive SWAP measurement is performed pairwise between corresponding qubits of the two replicas. Each pair is rotated from the Bell basis to the computational basis using ${\rm CZ}$ followed by $H$ gates and then measured. The single-shot global SWAP estimator is constructed from the resulting bit strings according to Eq.~\ref{eq:main_swap_single_shot}. The ratio of the string-inserted and non-inserted SWAP averages gives the normalized R\'enyi-2 correlator $C_{i,k}^{(2)}=N_{i,k}/D$.
}
    \label{fig:circuit}
\end{figure*}

\subsection{Two-replica R\'enyi-2 string diagnostic of ASPT order}
\label{sec:main_two_replica}

The single-copy measurements above show that the conventional cluster-string
order is suppressed by decoherence and therefore cannot, by itself, serve as a
reliable diagnostic of the ASPT  state. We therefore move to a
density-matrix-level characterization using a two-replica R\'enyi-2 string
diagnostic
\cite{maAverageSymmetryProtected2023,maTopologicalPhasesAverage2025,
maSymmetryProtectedTopological2025,xueTensorNetworkFormulation2024,
guoLocallyPurifiedDensity2025}. Operationally, this requires two independently
prepared copies of the same mixed state. The two replicas, labeled $A$ and
$B$, are generated using cluster-state preparation and
ancilla-assisted dephasing circuits. Each
replica therefore realizes the same ensemble-averaged density matrix
$\rho_{\rm dec}$. In this setup, we characterize the mixed-state string order using the normalized
R\'enyi-2 correlator
\begin{equation}
C_{i,k}^{(2)}
=
\frac{
\Tr\!\left[
\rho_{\rm dec}\mathcal{O}_{i,k}
\rho_{\rm dec}\mathcal{O}_{i,k}
\right]
}{
\Tr\!\left(\rho_{\rm dec}^{2}\right)
}
\equiv
\frac{N_{i,k}}{D},
\label{eq:main_swap_ratio}
\end{equation}
where $\mathcal{O}_{i,k}$ is the cluster string operator defined in
Eq.~\ref{eq:main_cluster_string} (see Appendix.~\ref{sec:methods_doubled_string} for more details).

The complete workflow is summarized in Fig.~\ref{fig:circuit}.
We implement the two-replica protocol on the two-dimensional IBM Nighthawk
processor, whose connectivity allows the two system copies to be embedded in
close proximity while attaching an ancilla qubit to each dephased even-sublattice
site [Fig.~\ref{fig:circuit}(a)]. Using this layout, the two cluster states are
prepared independently according to Eq.~\ref{eq:main_cluster_preparation}, as
shown in Step~1 of Fig.~\ref{fig:circuit}(b).

We next express $C_{i,k}^{(2)}$ as a ratio of quantities obtained from
destructive SWAP measurements on the two replicas. We first consider the
denominator, which is the second R\'enyi purity [Step~2 in
Fig.~\ref{fig:circuit}(b)],
\begin{equation}
D
=
\Tr(\rho_{\rm dec}^{2})
=
\Tr\!\left[
(\rho_{\rm dec}^{A}\otimes\rho_{\rm dec}^{B})S_{AB}
\right],
\label{eq:main_swap_denominator}
\end{equation}
where
$S_{AB}=\prod_{j=1}^{L}S_{j_Aj_B}$
is the global SWAP operator between replicas $A$ and $B$. To evaluate this
quantity, we perform a Bell-basis measurement on each corresponding replica
pair $(j_A,j_B)$ using the gate sequence shown in Step~3 of
Fig.~\ref{fig:circuit}(b), followed by destructive computational-basis readout.
Denoting the measured bits by $(a_j,b_j)\in\{0,1\}^2$, the local SWAP
eigenvalue for rung $j$ is
\begin{equation}
s_j=(-1)^{a_jb_j}.
\end{equation}
The resulting single-shot estimator of the global SWAP is therefore
\begin{equation}
S_{AB}^{\rm shot}
=
\prod_{j=1}^{L}s_j
=
(-1)^{\sum_{j=1}^{L}a_jb_j}.
\label{eq:main_swap_single_shot}
\end{equation}
Averaging this estimator over repeated shots yields $\left\langle S_{AB}^{\rm shot}\right\rangle
=
\langle S_{AB}\rangle
=
D
=
\Tr(\rho_{\rm dec}^{2}).$

The numerator is obtained in an analogous way, but with the string operator
$\mathcal{O}_{i,k}$ applied to one replica before the destructive SWAP
measurement.   Since $\mathcal{O}_{i,k}$ is
Hermitian and unitary, we have
\begin{equation}
\begin{aligned}
N_{i,k}
&=
\Tr\!\left[
\rho_{\rm dec}\mathcal{O}_{i,k}
\rho_{\rm dec}\mathcal{O}_{i,k}
\right]
\\
&=
\Tr\!\left[
\left(
\mathcal{O}_{i,k}^{A}\rho_{\rm dec}^{A}\mathcal{O}_{i,k}^{A}
\otimes
\rho_{\rm dec}^{B}
\right)
S_{AB}
\right].
\end{aligned}
\label{eq:main_swap_numerator}
\end{equation}
Operationally, the second line gives the measurement protocol directly:
$\mathcal{O}_{i,k}$ is applied only to replica $A$ immediately before the
same destructive SWAP readout used for the denominator [see Step 2 in Fig.~\ref{fig:circuit} (b)]. Thus, the numerator
and denominator require the same two-replica measurement circuit, differing
only by the insertion of the string operator on one replica.

\begin{figure*}
    \centering
    \includegraphics[width=0.7\linewidth]{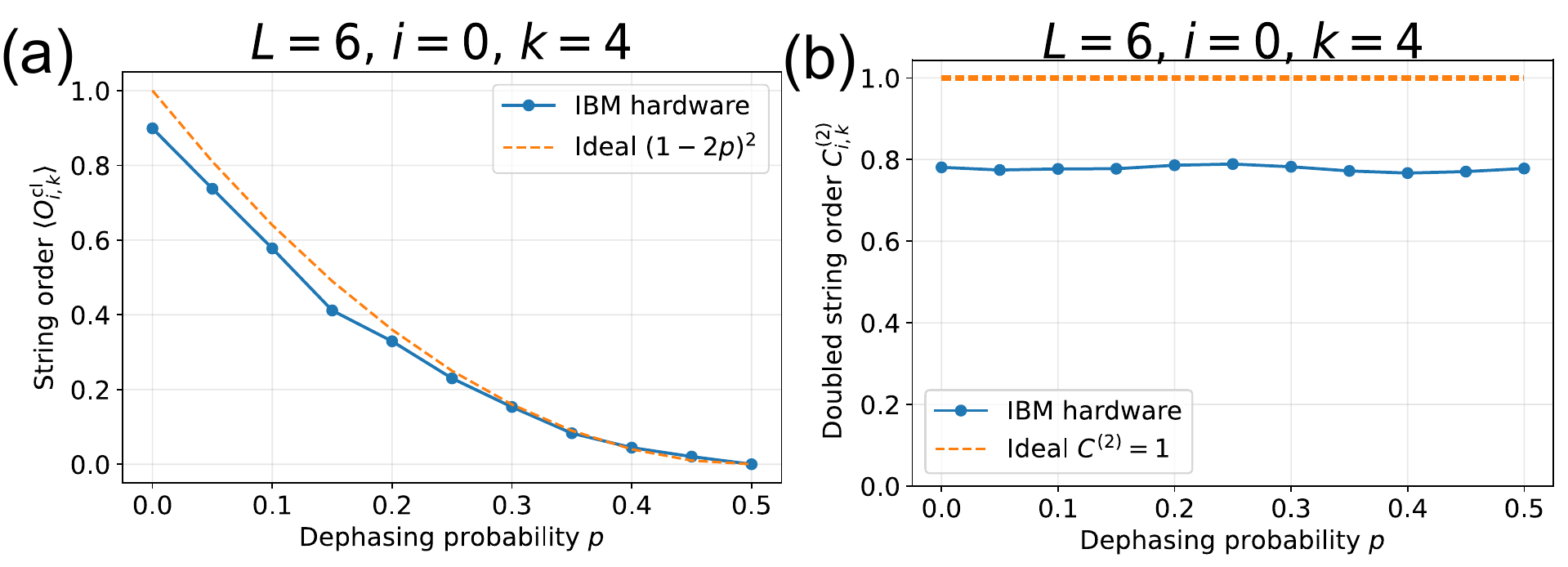}
    \caption{
Comparison between ordinary and doubled string diagnostics.
(a), Single-copy string correlator for the decohered $L=6$ cluster chain. The measured signal decreases with increasing dephasing probability $p$, consistent with the ideal decay in Eq.~\eqref{eq:main_string_ij_decay}. The corresponding ancilla-assisted dephasing circuit is shown in Fig.~\ref{fig:device}(b).
(b) Normalized R\'enyi-2 string correlator $C_{i,k}^{(2)}$ for the decohered cluster chain. In contrast to the single-copy signal, $C_{i,k}^{(2)}$ remains nearly independent of $p$ and close to the ideal value $C_{i,k}^{(2)}=1$. The two-replica measurement circuit is shown in Fig.~\ref{fig:circuit}, with each replica containing $L=6$ system qubits. At $p=0$, the system--ancilla CZ layer is retained in the hardware circuit, although it produces no dephasing in the ideal circuit; the associated gate therefore leads to a residual deviation. The contrasting behaviors here demonstrate that trajectory-dependent sign cancellation suppresses the conventional string correlator, whereas the two-replica diagnostic retains the string-order signature encoded in the ASPT density matrix.
}
    \label{fig:res1}
\end{figure*}

We now compare the single-copy and two-replica diagnostics for the decohered
cluster chain in Fig.~\ref{fig:res1}, using the representative $L=6$ circuit
shown in Fig.~\ref{fig:circuit}(a). The ideal single-copy string signal
follows
\begin{equation}
\langle \mathcal{O}_{i,k}\rangle_{\rm dec}
=
(1-2p)^{N_{\rm dep}(i,k)},
\label{eq:main_string_ik_decay}
\end{equation}
where $N_{\rm dep}(i,k)$ counts the dephased even-sublattice sites on which
$\mathcal{O}_{i,k}$ has $X$ support. For the string $\mathcal{O}_{i,k}$
considered here,  we choose the endpoints
such that two of its $X$ operators lie on dephased even-sublattice sites, so that $N_{\rm dep}(i,k)=2$. The normalized R\'enyi-2 string correlator $C_{i,k}^{(2)}$ exhibits a qualitatively different response. Each dephasing trajectory is then a locally $Z$-dressed cluster state, $|\psi_{\bm{\eta}}\rangle=Z(\bm{\eta})|\psi_{\rm cl}\rangle$, satisfying $\mathcal{O}_{i,k}|\psi_{\bm{\eta}}\rangle = s_{\bm{\eta}}|\psi_{\bm{\eta}}\rangle,~s_{\bm{\eta}}=\pm1$. The ordinary single-copy string correlator averages these trajectory-dependent signs and can therefore be suppressed by their cancellation. By contrast, the R\'enyi-2 diagnostic depends on $s_{\bm{\eta}}^2=1$. Since distinct dephasing sectors are orthogonal, we have
\begin{equation} 
C_{(i,k)~{\rm Ideal}}^{(2) } = \frac{ \sum_{\bm{\eta}}P(\bm{\eta})^2s_{\bm{\eta}}^2 }{ \sum_{\bm{\eta}}P(\bm{\eta})^2 } = 1 . \label{eq:main_doubled_string_fixed_point} 
\end{equation}
 Thus, the two-replica diagnostic eliminates the trajectory-sign cancellation that suppresses the conventional single-copy string correlator and retains the hidden string order of the ASPT  state.

Figs.~\ref{fig:res1}(a,b) compare the conventional single-copy string
correlator with the normalized R\'enyi-2 string diagnostic as the dephasing
probability $p$ is increased. While the measured
$\langle\mathcal{O}_{i,k}\rangle_{\rm dec}$ as shown in Fig.~\ref{fig:res1} (a) is progressively suppressed, the
measured $C_{i,k}^{(2)}$ remains nearly unchanged and close to the ideal value
$C_{i,k}^{(2)}=1$ predicted by
Eq.~\ref{eq:main_doubled_string_fixed_point} [Fig.~\ref{fig:res1} (b)]. This contrasting behavior
provides the key signature of the ASPT character of the
prepared mixed state: the persistence
of $C_{i,k}^{(2)}$ shows that the corresponding string structure remains encoded
in the density matrix and is retained by the two-replica diagnostic despite
the engineered dephasing. More details are provided in Appendix.~\ref{sec:methods_doubled_string}.

\subsection{Measurement of ASPT degeneracy in density-matrix spectra}
We next move beyond the above observable-level measurement. As a complementary
density-matrix-level characterization, we reconstruct the half-chain entanglement
spectrum of the coherent and dephasing-dressed cluster state for an $L=8$ system
\cite{liHaldanePhaseS2008,pollmannEntanglementSpectrumTopological2010,
pollmannDetectionSymmetryProtected2012,pollmannSymmetryProtectionTopological2012,
sonTopologicalOrderCluster2011,elseSymmetryProtectedPhases2012,Yu2024PRL,Zhong2025PRB,Guo2026PRR}. Unlike the
two-replica R\'enyi-2 diagnostic, this characterization is obtained by
reconstructing $\rho$ from repeated measurements on single prepared copies and
therefore does not require the simultaneous preparation of two replicas. From
the reconstructed density matrix, we trace out one half of the chain, $\rho_A=\Tr_{\bar A}\rho,$
where $A$ denotes one half of the chain and $\bar A$ its complement. Denoting
the eigenvalues of $\rho_A$ by $\lambda_\alpha$, the corresponding spectrum is
\begin{equation}
\xi_\alpha=-\log\lambda_\alpha .
\label{eq:main_entanglement_spectrum}
\end{equation}

For the ideal cluster state, this spectrum exhibits the characteristic twofold
degeneracy associated with the SPT edge structure
\cite{liHaldanePhaseS2008,pollmannEntanglementSpectrumTopological2010,
pollmannDetectionSymmetryProtected2012,pollmannSymmetryProtectionTopological2012}. The engineered dephasing preserves this structure at the trajectory level because
each trajectory is a locally $Z$-dressed cluster state, $|\psi_{\bm{\eta}}\rangle=Z(\bm{\eta})|\psi_{\rm cl}\rangle$. Across the bipartition, $Z(\bm{\eta})
=
Z_A(\bm{\eta}_A)\otimes Z_{\bar A}(\bm{\eta}_{\bar A}),$
so the dressing acts as independent local unitaries on the two sides of the
cut. These operations can rotate the Schmidt vectors but do not change the
Schmidt values, and hence each dephasing trajectory retains the twofold
degeneracy of the ideal cluster state
\cite{liHaldanePhaseS2008,pollmannEntanglementSpectrumTopological2010,
pollmannDetectionSymmetryProtected2012,pollmannSymmetryProtectionTopological2012} (see Appendix~\ref{sec:methods_entanglement_spectrum} for details)

\begin{figure*}
    \centering
    \includegraphics[width=0.7\linewidth]{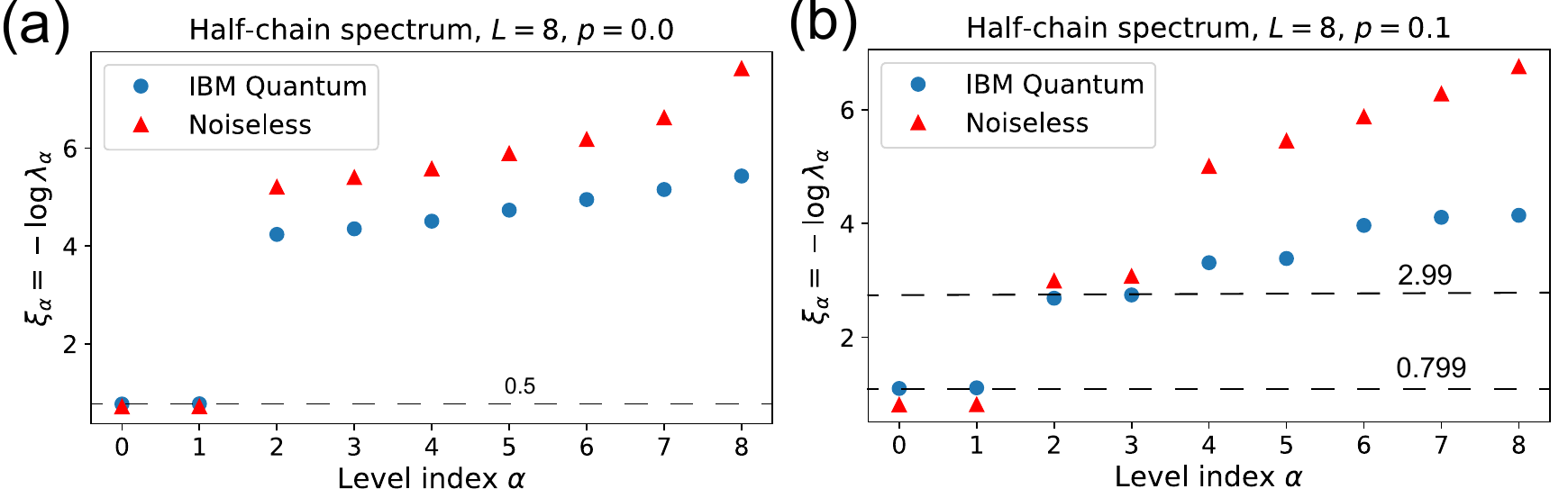}
    \caption{
Half-chain entanglement spectrum reconstructed from the measured density matrix. 
For an $L=8$ chain, we trace out one half of the system to obtain the reduced density matrix, and then compute the entanglement levels from Eq.~\ref{eq:main_entanglement_spectrum}.
Blue circles denote IBM Quantum data, and red triangles denote the noiseless reference.
(a) Measured spectrum $\xi_\alpha=-\log\lambda_\alpha$ for an $L=8$ cluster state at $p=0$. The two lowest levels form the characteristic nearly degenerate pair associated with the cluster-state edge sector.
(b) Corresponding spectrum at dephasing probability $p=0.1$. Engineered dephasing redistributes spectral weight into an additional reduced-stabilizer sector while preserving the pairwise structure of the low-lying spectrum. Dashed lines indicate the analytical ideal levels, $\xi_{1,2}\simeq0.799$ and $\xi_{3,4}\simeq2.996$. Deviations of the higher levels from the noiseless reference reflect the increased sensitivity of the small-eigenvalue sector to hardware imperfections and finite sampling.
See Appendix.~\ref{sec:methods_entanglement_spectrum} for a more detailed explanation.
}
    \label{fig:res4}
\end{figure*}

After the trajectory record is discarded in our ancilla-based circuit, the
reduced density matrix becomes a classical mixture of locally dressed reduced
density matrices,
\begin{equation}
\rho_A^{\rm dec}
=
\sum_{\bm{\eta}_A}
P_A(\bm{\eta}_A)
Z_A(\bm{\eta}_A)
\rho_A^{\rm cl}
Z_A(\bm{\eta}_A).
\label{eq:main_reduced_density_trajectory_mixture}
\end{equation}
The effect of the dephasing trajectories is to redistribute weight among
different reduced-stabilizer, or bulk-syndrome, sectors of subsystem $A$
\cite{pollmannEntanglementSpectrumTopological2010,
pollmannDetectionSymmetryProtected2012,pollmannSymmetryProtectionTopological2012,
maAverageSymmetryProtected2023,maTopologicalPhasesAverage2025}.
Crucially, the local $Z$ insertions do not resolve the virtual edge doublet. Each populated syndrome sector therefore retains the same twofold edge degeneracy, giving the schematic block structure \begin{equation} \rho_A^{\rm dec} = \sum_{\mu} w_\mu\, \rho_{A,{\rm bulk}}^{(\mu)} \otimes \frac{\mathbb{I}_{\rm edge}}{2}, \label{eq:main_dephased_reduced_block_structure} \end{equation} where $\mu$ labels the dephasing-induced bulk sectors and $w_\mu$ denotes their classical weights \cite{liHaldanePhaseS2008,pollmannEntanglementSpectrumTopological2010, pollmannDetectionSymmetryProtected2012,pollmannSymmetryProtectionTopological2012, maSymmetryProtectedTopological2025,xueTensorNetworkFormulation2024, guoLocallyPurifiedDensity2025}. Thus, although dephasing modifies the spectral weights of $\rho_A$, the characteristic pairwise edge degeneracy remains visible in the reconstructed density matrix. This provides evidence that the prepared mixed state retains the SPT edge structure after ensemble averaging, despite the decay of the conventional single-copy string signal. Further details of the block structure are given in Appendix~\ref{sec:methods_entanglement_spectrum}.

Figure~\ref{fig:res4} compares the measured half-chain spectrum with the
corresponding ideal predictions at $p=0$ and $p=0.1$. At $p=0$, the cluster
state exhibits the expected twofold-degenerate spectrum associated with the
projective edge structure of the one-dimensional cluster SPT. For the $L=8$
bipartition considered here, the ideal reduced density matrix has two equal
nonzero eigenvalues, $\lambda_{1,2}=1/2$, corresponding to
$\xi_{1,2}=\ln 2$. The measured spectrum closely reproduces this leading
twofold pairing.

When the engineered dephasing is introduced at $p=0.1$, the reduced density
matrix acquires an additional stabilizer-syndrome sector. Importantly, the
dephasing does not simply broaden or uniformly suppress the original spectrum;
instead, it redistributes spectral weight between distinct reduced-stabilizer
sectors while preserving the two-dimensional edge subspace within each sector.
For the ideal $L=8$ half-chain, this gives
$\lambda_{1,2}=0.45$ and $\lambda_{3,4}=0.05$, or equivalently
$\xi_{1,2}\simeq0.799$ and $\xi_{3,4}\simeq2.996$
[see Appendix~\ref{sec:methods_entanglement_spectrum}]. Thus, the appearance of
additional entanglement levels under dephasing is itself expected and reflects
the redistribution of probability among dephasing-induced syndrome sectors,
rather than a lifting of the characteristic edge degeneracy.

The measured spectrum in Fig.~\ref{fig:res4}(b) follows this structure: the
dominant levels remain nearly degenerate, while a higher-lying paired sector
emerges at the scale predicted by the ideal dephasing model. The
ancilla-assisted implementation introduces small deviations, most visibly in
these higher levels, which are more sensitive to finite sampling and hardware
imperfections because they carry substantially smaller spectral weight.
Nevertheless, the pairwise organization of the spectrum remains clearly
resolved. This provides a complementary density-matrix-level signature that
the engineered dephasing changes the bulk syndrome content of the reduced state
without lifting the characteristic twofold edge-sector degeneracy.

\section{Discussion}
\label{sec:discussion}

In this work, we realize a decoherence-induced averaged SPT state on
programmable quantum processors by engineering sublattice-selective Pauli-$Z$
dephasing of a one-dimensional cluster SPT. The resulting channel converts the even-sublattice symmetry from a strong symmetry to an average symmetry while leaving the odd-sublattice symmetry strong. Our measurements reveal that the symmetry
charge and conventional single-copy string correlator decay under dephasing,
whereas the two-replica R\'enyi-2 string correlator, defined at the
density-matrix level, remains nontrivial. Moreover, the reconstructed
half-chain spectrum further shows that dephasing redistributes spectral weight
among reduced-stabilizer sectors while preserving the characteristic twofold
pairing.

Our results establish engineered decoherence as a controllable route to
mixed-state topological order on gate-based quantum hardware. More broadly,
this framework can be extended to two-dimensional cluster SPT states exhibiting
decoherence-driven topological transitions
\cite{guo2024twodimensional}, interacting SPT states away from the exactly
solvable cluster limit \cite{leeSymmetryProtectedTopological2025}, and intrinsically mixed-state
topological phases generated from topologically ordered states such as the
toric code \cite{wang2025intrinsic,sang2024mixed}. Spatially correlated decoherence provides another natural direction for exploring how the
structure of the environment controls the stability and transition of
mixed-state topological order \cite{chirame2025stable}.

\begin{acknowledgments}
We acknowledge the use of IBM Quantum services for this work.  The views expressed are those of the authors and do not reflect the official policy or position of IBM or the IBM Quantum team.  All data and code for this work are available from the corresponding authors upon reasonable request.  This work is supported by the Singapore Ministry of Education Academic Research Fund Tier-II Grant (MOE-T2EP50224-0007). X.-J. Yu was supported by the National Natural Science Foundation of China (Grant No.12405034) and a start-up grant from Eastern Institute of Technology, Ningbo. 
\end{acknowledgments}

\bibliography{ref}
\newpage
\onecolumngrid
\flushbottom
\newpage
\appendix
\setcounter{equation}{0}
\setcounter{figure}{0}
\setcounter{table}{0}
\setcounter{section}{0}
\renewcommand{\theequation}{S\arabic{equation}}
\renewcommand{\thefigure}{S\arabic{figure}}
\renewcommand{\thesection}{S\arabic{section}}
\renewcommand{\thepage}{S\arabic{page}}

\section*{Appendix}

\section{Cluster-state conventions and boundary terms}
\label{sec:methods_cluster_conventions}

We summarize the cluster-state conventions used in the main text and clarify
the role of boundary terms. In the bulk of a one-dimensional chain, the cluster
stabilizers are
\begin{equation}
K_j
=
Z_{j-1}X_jZ_{j+1}.
\label{eq:methods_bulk_stabilizer}
\end{equation}
For an open chain, two closely related conventions are commonly used. In the
graph-state convention adopted for state preparation in this work, the boundary
stabilizers
\begin{equation}
K_1=X_1Z_2,
\qquad
K_L=Z_{L-1}X_L
\label{eq:methods_boundary_stabilizers}
\end{equation}
are included together with Eq.~\eqref{eq:methods_bulk_stabilizer}. The resulting
cluster graph state is the unique simultaneous $+1$ eigenstate of the complete
stabilizer set
\cite{raussendorfOneWayQuantum2001,heinEntanglementGraphStates2004,
heinEntanglementGraphStatesApplications2006},
\begin{equation}
K_j|\psi_{\rm cl}\rangle
=
|\psi_{\rm cl}\rangle,
\qquad
j=1,\ldots,L.
\label{eq:methods_stabilizer_condition}
\end{equation}
Alternatively, in the symmetry-preserving open-chain convention, only the bulk
stabilizers $K_j$ with $1<j<L$ are included. The two unconstrained boundary
degrees of freedom then generate the characteristic edge-state degeneracy of
the cluster SPT phase. Thus, the first convention specifies the 
state prepared in our circuits.

The graph-state preparation circuit for the open-chain cluster state is
\begin{equation}
\ket{\psi_{\rm cl}}
=
U_{\rm CZ}^{\rm odd}
U_{\rm CZ}^{\rm even}
\ket{+}^{\otimes L},
\label{eq:methods_cluster_circuit}
\end{equation}
with
\begin{equation}
U_{\rm CZ}^{\rm even}
=
\prod_{\substack{j\ {\rm even}\\1\leq j<L}}
{\rm CZ}_{j,j+1},
\qquad
U_{\rm CZ}^{\rm odd}
=
\prod_{\substack{j\ {\rm odd}\\1\leq j<L}}
{\rm CZ}_{j,j+1}.
\label{eq:methods_cz_sublayers}
\end{equation}
Gates within the same sublayer act on disjoint pairs and can therefore be
executed in parallel on a nearest-neighbour device.

\section{Ancilla implementation of dephasing channels}
\label{sec:methods_ancilla_dephasing}

We provide additional details on the implementation of dephasing channels in quantum circuits. For each even-sublattice system qubit $q_j$, an ancilla $a_j$ is initialized in
$|0\rangle_{a_j}$ and prepared as
\begin{equation}
R_y(2\theta)|0\rangle_{a_j}
=
\sqrt{1-p}\,|0\rangle_{a_j}
+
\sqrt{p}\,|1\rangle_{a_j},
\qquad
p=\sin^2\theta.
\label{eq:methods_ancilla_rotation}
\end{equation}
The ancilla is then coupled to the system qubit through a controlled-$Z$
operation,
\begin{equation}
U_{a_jq_j}
=
|0\rangle\langle0|_{a_j}\otimes I_j
+
|1\rangle\langle1|_{a_j}\otimes Z_j.
\label{eq:methods_controlled_z_noise}
\end{equation}
Measuring the ancilla in the computational basis gives the two Kraus operators
\cite{krausStatesEffectsOperations1983,
nielsenQuantumComputationQuantum2010,
wildeQuantumInformationTheory2017}
\begin{equation}
M_{j,0}
=
\sqrt{1-p}\,I_j,
\qquad
M_{j,1}
=
\sqrt{p}\,Z_j.
\label{eq:methods_dephasing_kraus}
\end{equation}
Discarding the ancilla outcome therefore produces the local dephasing channel
\begin{equation}
\mathcal{E}_j^Z(\rho)
=
\sum_{\eta_j=0,1}
M_{j,\eta_j}\rho M_{j,\eta_j}^\dagger
=
(1-p)\rho+pZ_j\rho Z_j.
\label{eq:methods_dephasing_channel}
\end{equation}
Applying the construction independently to all even-sublattice sites gives
\begin{equation}
\mathcal{E}_{\rm even}^Z
=
\prod_{j\in{\rm even}}\mathcal{E}_j^Z.
\label{eq:methods_even_channel}
\end{equation}

If the ancilla measurement record is retained, the same circuit admits a
trajectory-resolved description
\cite{maAverageSymmetryProtected2023,maTopologicalPhasesAverage2025,
maSymmetryProtectedTopological2025,zhangQuantumCommunicationMixed2025,
luNonequilibriumTopologicalResponse2026}. We denote the outcome at even site
$j$ by $\eta_j\in\{0,1\}$, corresponding respectively to the identity and
$Z_j$ branches. For the complete record
$\bm{\eta}=\{\eta_j\}_{j\in{\rm even}}$, the trajectory operator and its
probability are
\begin{equation}
Z(\bm{\eta})
=
\prod_{j\in{\rm even}}Z_j^{\eta_j},
\qquad
P(\bm{\eta})
=
\prod_{j\in{\rm even}}
p^{\eta_j}(1-p)^{1-\eta_j}.
\label{eq:methods_record_probability}
\end{equation}
The corresponding conditional system state is
\begin{equation}
|\psi_{\bm{\eta}}\rangle
=
Z(\bm{\eta})|\psi_{\rm cl}\rangle,
\label{eq:methods_trajectory_state}
\end{equation}
and averaging over all measurement records yields
\begin{equation}
\rho_{\rm dec}
=
\sum_{\bm{\eta}}
P(\bm{\eta})
|\psi_{\bm{\eta}}\rangle
\langle\psi_{\bm{\eta}}|,
\label{eq:methods_trajectory_ensemble}
\end{equation}
which is equivalent to applying
$\mathcal{E}_{\rm even}^Z$ to the initial cluster-state density matrix.

\section{Detailed structure of the averaged cluster SPT}
\label{sec:supp_z_dephasing_aspt_justification}
In this section, we justify the use of local $Z$ dephasing in the decohered
cluster-chain protocol and explain how the resulting mixed state realizes the
strong--average symmetry structure of an ASPT  phase. The decoherence
introduced here is not treated as a generic uncontrolled error. Instead, it is
an engineered local quantum channel with a definite symmetry action: the
odd-sublattice symmetry remains strong, whereas the even-sublattice symmetry
is reduced from a strong symmetry to a weak, or average, symmetry after
ensemble averaging.

The local channel applied to each even-sublattice site is
\begin{equation}
\mathcal{E}_j^Z(\rho)
=
(1-p)\rho+pZ_j\rho Z_j,
\label{eq:supp_local_z_dephasing}
\end{equation}
where $p$ is the probability of the local $Z_j$ branch. This is the standard
single-site Pauli-$Z$ dephasing channel
\cite{krausStatesEffectsOperations1983,nielsenQuantumComputationQuantum2010,
wildeQuantumInformationTheory2017}. For a single-qubit density matrix
\begin{equation}
\rho=
\begin{pmatrix}
\rho_{00} & \rho_{01}\\
\rho_{10} & \rho_{11}
\end{pmatrix},
\end{equation}
the channel gives
\begin{equation}
\mathcal{E}^Z(\rho)
=
\begin{pmatrix}
\rho_{00} & (1-2p)\rho_{01}\\
(1-2p)\rho_{10} & \rho_{11}
\end{pmatrix}.
\label{eq:supp_single_qubit_dephasing_matrix}
\end{equation}
Thus, the computational-basis populations are unchanged, while the
off-diagonal coherences are multiplied by $1-2p$. In the range
$0\leq p\leq1/2$, $p$ therefore provides a monotonic measure of the dephasing
strength: $p=0$ corresponds to the identity channel, whereas $p=1/2$ gives
complete dephasing in the $Z$ basis.

Applying the channel independently to all even-sublattice sites gives
\begin{equation}
\rho_{\rm dec}
=
\left(
\prod_{j\in{\rm even}}
\mathcal{E}_j^Z
\right)
(\rho_{\rm cl}),
\label{eq:supp_even_dephased_state}
\end{equation}
where $\rho_{\rm cl}=|\psi_{\rm cl}\rangle\langle\psi_{\rm cl}|$ is the
initial cluster-state density matrix. Equivalently, the resulting state can be
written as the trajectory ensemble
\begin{equation}
\rho_{\rm dec}
=
\sum_{\bm{\eta}}
P(\bm{\eta})
Z(\bm{\eta})
\rho_{\rm cl}
Z(\bm{\eta}),
\label{eq:supp_dephasing_ensemble}
\end{equation}
with
\begin{equation}
Z(\bm{\eta})
=
\prod_{j\in{\rm even}}Z_j^{\eta_j},
\qquad
P(\bm{\eta})
=
\prod_{j\in{\rm even}}
p^{\eta_j}(1-p)^{1-\eta_j},
\label{eq:supp_dephasing_probability}
\end{equation}
where $\eta_j\in\{0,1\}$ labels the identity and $Z_j$ branches,
respectively.

This form makes the symmetry action transparent. Since the dephasing operators
act only on the even sublattice, they commute with the odd-sublattice generator
$U_{\rm odd}$, so the odd symmetry remains strong. By contrast, the local
$Z_j$ operators anticommute with the corresponding $X_j$ factors in
$U_{\rm even}$, so individual trajectories generally carry different
even-sublattice symmetry charges. After averaging over the trajectories,
however, the density matrix remains invariant under conjugation by
$U_{\rm even}$,
\begin{equation}
U_{\rm even}\rho_{\rm dec}U_{\rm even}^{\dagger}
=
\rho_{\rm dec},
\label{eq:supp_even_average_symmetry}
\end{equation}
while it no longer needs to satisfy the stronger condition
$U_{\rm even}\rho_{\rm dec}=\rho_{\rm dec}$. The engineered dephasing
therefore converts the even-sublattice symmetry from strong to average while
preserving the odd-sublattice symmetry as strong, yielding the hybrid symmetry
structure
\begin{equation}
\mathbb{Z}_2^{{\rm odd},{\rm strong}}
\times
\mathbb{Z}_2^{{\rm even},{\rm average}},
\label{eq:supp_strong_average_structure}
\end{equation}
which underlies the decoherence-induced averaged cluster SPT construction used
in this work
\cite{maAverageSymmetryProtected2023,maTopologicalPhasesAverage2025,
maSymmetryProtectedTopological2025,xueTensorNetworkFormulation2024,
guoLocallyPurifiedDensity2025}.

We now explain why this channel produces the desired averaged cluster-chain SPT.  The fixed-point cluster state is protected by the two $\mathbb{Z}_2$ symmetry generators
\begin{equation}
U_{\rm odd}=\prod_{j\in{\rm odd}}X_j,
\qquad
U_{\rm even}=\prod_{j\in{\rm even}}X_j .
\label{eq:supp_cluster_symmetry_generators}
\end{equation}
We assume a boundary convention for which the reference cluster state lies in the $+1$ sector of both symmetry generators, for example periodic boundary conditions or an open-chain fixed-point state with the boundary degrees of freedom fixed.  Its density matrix
\begin{equation}
\rho_{\rm cl}=\ket{\psi_{\rm cl}}\bra{\psi_{\rm cl}}
\label{eq:supp_pure_cluster_density_matrix}
\end{equation}
then obeys
\begin{equation}
U_{\alpha}\rho_{\rm cl}=\rho_{\rm cl}U_{\alpha}=\rho_{\rm cl},
\qquad
\alpha\in\{{\rm odd},{\rm even}\}.
\label{eq:supp_pure_strong_symmetry}
\end{equation}
Here, a strong symmetry means that the density matrix is fixed by the symmetry operator when the latter acts from either the left or the right.

In the decohered protocol, $Z$-dephasing is applied only to the even sublattice:
\begin{equation}
\rho_{\rm dec}=\prod_{j\in{\rm even}}\mathcal{E}_j^Z(\rho_{\rm cl}).
\label{eq:supp_even_sublattice_dephased_state}
\end{equation}
Equivalently, the mixed state can be written as an ensemble average
\begin{equation}
\rho_{\rm dec}=\sum_{\boldsymbol{\eta}}P(\boldsymbol{\eta})Z(\boldsymbol{\eta})\rho_{\rm cl}Z(\boldsymbol{\eta}),
\label{eq:supp_dephased_ensemble}
\end{equation}
where
\begin{equation}
Z(\boldsymbol{\eta})=\prod_{j\in{\rm even}}Z_j^{\eta_j},
\qquad
\eta_j\in\{0,1\},
\label{eq:supp_z_eta}
\end{equation}
and
\begin{equation}
P(\boldsymbol{\eta})=\prod_{j\in{\rm even}}p^{\eta_j}(1-p)^{1-\eta_j}.
\label{eq:supp_eta_probability}
\end{equation}
The binary string $\boldsymbol{\eta}$ labels the Kraus trajectory of the engineered dephasing channel.  In an ancilla-assisted implementation, the same label can be identified with the corresponding recorded measurement branch, and Eq.~\eqref{eq:supp_dephased_ensemble} is obtained after averaging over the measurement record.

The selectivity of the protocol follows from the sublattice structure of the noise.  For $j\in{\rm even}$, the operator $Z_j$ commutes with $U_{\rm odd}$, because $U_{\rm odd}$ acts only on odd sites:
\begin{equation}
[U_{\rm odd},Z_j]=0,
\qquad
j\in{\rm even}.
\label{eq:supp_odd_commutes_with_z_noise}
\end{equation}
Therefore
\begin{equation}
U_{\rm odd}Z(\boldsymbol{\eta})=Z(\boldsymbol{\eta})U_{\rm odd}.
\label{eq:supp_odd_commutes_with_z_eta}
\end{equation}
Using Eq.~\eqref{eq:supp_pure_strong_symmetry}, one obtains
\begin{align}
U_{\rm odd}\rho_{\rm dec}
&=\sum_{\boldsymbol{\eta}}P(\boldsymbol{\eta})U_{\rm odd}Z(\boldsymbol{\eta})\rho_{\rm cl}Z(\boldsymbol{\eta})
\nonumber\\
&=\sum_{\boldsymbol{\eta}}P(\boldsymbol{\eta})Z(\boldsymbol{\eta})U_{\rm odd}\rho_{\rm cl}Z(\boldsymbol{\eta})
\nonumber\\
&=\sum_{\boldsymbol{\eta}}P(\boldsymbol{\eta})Z(\boldsymbol{\eta})\rho_{\rm cl}Z(\boldsymbol{\eta})
=\rho_{\rm dec}.
\label{eq:supp_odd_left_action}
\end{align}
The same argument for right multiplication gives
\begin{equation}
\rho_{\rm dec}U_{\rm odd}=\rho_{\rm dec}.
\label{eq:supp_odd_right_action}
\end{equation}
Hence the odd symmetry remains strong after decoherence:
\begin{equation}
U_{\rm odd}\rho_{\rm dec}=\rho_{\rm dec}U_{\rm odd}=\rho_{\rm dec}.
\label{eq:supp_odd_strong_after_dephasing}
\end{equation}

The even symmetry behaves differently.  Since $U_{\rm even}$ contains $X_j$ on every even site, while the dephasing channel inserts $Z_j$ on even sites, one has
\begin{equation}
X_jZ_j=-Z_jX_j,
\qquad
j\in{\rm even}.
\label{eq:supp_xz_anticommutation}
\end{equation}
Therefore,
\begin{equation}
U_{\rm even}Z(\boldsymbol{\eta})=(-1)^{|\boldsymbol{\eta}|}Z(\boldsymbol{\eta})U_{\rm even},
\qquad
|\boldsymbol{\eta}|=\sum_{j\in{\rm even}}\eta_j .
\label{eq:supp_even_anticommutes_z_eta}
\end{equation}
Acting from the left gives
\begin{align}
U_{\rm even}\rho_{\rm dec}
&=\sum_{\boldsymbol{\eta}}P(\boldsymbol{\eta})U_{\rm even}Z(\boldsymbol{\eta})\rho_{\rm cl}Z(\boldsymbol{\eta})
\nonumber\\
&=\sum_{\boldsymbol{\eta}}P(\boldsymbol{\eta})(-1)^{|\boldsymbol{\eta}|}Z(\boldsymbol{\eta})U_{\rm even}\rho_{\rm cl}Z(\boldsymbol{\eta})
\nonumber\\
&=\sum_{\boldsymbol{\eta}}P(\boldsymbol{\eta})(-1)^{|\boldsymbol{\eta}|}Z(\boldsymbol{\eta})\rho_{\rm cl}Z(\boldsymbol{\eta}).
\label{eq:supp_even_left_action}
\end{align}
For generic $p$, this is not equal to $\rho_{\rm dec}$.  Hence $U_{\rm even}$ is no longer a strong symmetry:
\begin{equation}
U_{\rm even}\rho_{\rm dec}\neq\rho_{\rm dec}.
\label{eq:supp_even_not_strong}
\end{equation}
Equivalently, the individual trajectory state $Z(\boldsymbol{\eta})\ket{\psi_{\rm cl}}$ carries a trajectory-dependent sign under the left action of $U_{\rm even}$.

Nevertheless, the even symmetry is preserved in the weak, or average, sense.  Under conjugation by $U_{\rm even}$, the signs from the left and right actions cancel:
\begin{align}
&U_{\rm even}Z(\boldsymbol{\eta})\rho_{\rm cl}Z(\boldsymbol{\eta})U_{\rm even}^{\dagger}
\nonumber\\
&\qquad=
(-1)^{|\boldsymbol{\eta}|}Z(\boldsymbol{\eta})U_{\rm even}\rho_{\rm cl}U_{\rm even}^{\dagger}Z(\boldsymbol{\eta})(-1)^{|\boldsymbol{\eta}|}
\nonumber\\
&\qquad=Z(\boldsymbol{\eta})\rho_{\rm cl}Z(\boldsymbol{\eta}) .
\label{eq:supp_even_conjugation_single_term}
\end{align}
After summing over all trajectories, this gives
\begin{equation}
U_{\rm even}\rho_{\rm dec}U_{\rm even}^{\dagger}=\rho_{\rm dec}.
\label{eq:supp_even_weak_symmetry}
\end{equation}
Thus the even generator is not a strong symmetry of the decohered density matrix, but it remains a weak symmetry of the ensemble-averaged mixed state.

The resulting mixed state therefore has the symmetry structure
\begin{equation}
\mathbb{Z}_2^{{\rm odd},{\rm strong}}\times
\mathbb{Z}_2^{{\rm even},{\rm weak}} .
\label{eq:supp_strong_weak_symmetry_structure}
\end{equation}
This is the symmetry structure characteristic of the decohered averaged cluster-chain SPT.

This also clarifies why the choice of even-sublattice $Z$-dephasing is selective.  If the noise commuted with both $U_{\rm odd}$ and $U_{\rm even}$, both symmetries would remain strong and no strong-to-weak conversion would occur.  Conversely, if the noise destroyed the conjugation invariance under $U_{\rm even}$, the ASPT  structure would be lost.  Even-sublattice $Z$-dephasing realizes the intermediate case required here: it preserves one $\mathbb{Z}_2$ symmetry strongly while converting the other into an average symmetry.

\section{Ordinary and doubled string diagnostics}
\label{sec:methods_doubled_string}

We provide the details behind the ordinary and doubled string diagnostics used in the main text.  

\subsection{Single-copy string diagnostic}
Let the endpoints $i$ and $k$ lie on the same sublattice, so that $k-i$ is
even. The corresponding cluster string operator is
\begin{equation}
\mathcal{O}_{i,k}
=
Z_i
\left(
\prod_{r=i+1,i+3,\ldots,k-1}
X_r
\right)
Z_k .
\label{eq:methods_string_operator}
\end{equation}
We denote the set of interior sites carrying $X$ support by $\mathcal{X}_{i,k}
=
\{i+1,i+3,\ldots,k-1\}.$

The ideal cluster state used throughout this work is prepared from
$|+\rangle^{\otimes L}$ by nearest-neighbor controlled-$Z$ gates,
\begin{equation}
|\psi_{\rm cl}\rangle
=
\left(
\prod_{j=1}^{L-1}{\rm CZ}_{j,j+1}
\right)
|+\rangle^{\otimes L},
\label{eq:methods_cluster_state_definition}
\end{equation}
and satisfies the bulk stabilizer conditions
$K_r|\psi_{\rm cl}\rangle=|\psi_{\rm cl}\rangle$, with
$K_r=Z_{r-1}X_rZ_{r+1}$. The string operator is the
product of the stabilizers on $\mathcal{X}_{i,k}$,
\begin{equation}
\mathcal{O}_{i,k}
=
\prod_{r\in\mathcal{X}_{i,k}}K_r .
\label{eq:methods_string_as_stabilizer_product}
\end{equation}
It therefore follows directly that
\begin{equation}
\mathcal{O}_{i,k}|\psi_{\rm cl}\rangle
=
|\psi_{\rm cl}\rangle.
\label{eq:methods_cluster_string_fixed_point}
\end{equation}

We first derive the single-copy decay under even-sublattice dephasing.  The local dephasing channel on an even site $j$ is
\begin{equation}
\mathcal{E}^{Z}_{j}(\rho)
=
(1-p)\rho+pZ_j\rho Z_j .
\label{eq:methods_local_dephasing_repeat}
\end{equation}
In the Heisenberg picture, we have for a Pauli operator $P$
\begin{equation}
\left(\mathcal{E}^{Z}_{j}\right)^\dagger(P)
=
(1-p)P+pZ_jPZ_j,
\label{eq:methods_channel_adjoint_general}
\end{equation}
 which simplifies to
\begin{equation}
\left(\mathcal{E}^{Z}_{j}\right)^\dagger(P)
=
\begin{cases}
P, & [P,Z_j]=0,\\
(1-2p)P, & \{P,Z_j\}=0.
\end{cases}
\label{eq:methods_channel_adjoint_pauli}
\end{equation}
For the engineered dephasing channel acting on the even sublattice, let
$\mathcal{D}_{\rm even}$ denote the set of dephased sites. Its adjoint action on
the string operator is
\begin{equation}
\left(\mathcal{E}^{Z}_{\rm even}\right)^\dagger
(\mathcal{O}_{i,k})
=
(1-2p)^{N_{\rm dep}(i,k)}
\mathcal{O}_{i,k},
\label{eq:methods_string_heisenberg_decay}
\end{equation}
where $N_{\rm dep}(i,k)
=
\left|
\mathcal{X}_{i,k}\cap\mathcal{D}_{\rm even}
\right|$
counts the dephased even-sublattice sites on which the string operator has
$X$ support. The string expectation value in the decohered state therefore
becomes
\begin{align}
\langle \mathcal{O}_{i,k}\rangle_{\rm dec}
&=
\Tr\!\left[
\mathcal{O}_{i,k}
\mathcal{E}^{Z}_{\rm even}(\rho_{\rm cl})
\right]
\nonumber\\
&=
\Tr\!\left[
\left(\mathcal{E}^{Z}_{\rm even}\right)^\dagger
(\mathcal{O}_{i,k})
\rho_{\rm cl}
\right]
\nonumber\\
&=
(1-2p)^{N_{\rm dep}(i,k)}
\Tr\!\left[
\mathcal{O}_{i,k}\rho_{\rm cl}
\right]
\nonumber\\
&=
(1-2p)^{N_{\rm dep}(i,k)}.
\label{eq:methods_single_string_decay}
\end{align}
Thus, the conventional string correlator is suppressed by averaging over
dephasing trajectories with different string signs.

\subsection{Double-copy string diagnostic}

We now turn to the doubled diagnostic.  The essential point is that the averaged-SPT order is encoded in the density matrix rather than in a linear single-copy expectation value.  We therefore use the Hilbert--Schmidt inner product
\begin{equation}
\langle\langle A|B\rangle\rangle
=
\Tr(A^\dagger B)
\label{eq:methods_hilbert_schmidt}
\end{equation}
and define the doubled string action by
\begin{equation}
\mathcal{O}^{(2)}_{i,k}:
|A\rangle\rangle
\mapsto
|\mathcal{O}_{i,k}A\mathcal{O}_{i,k}\rangle\rangle .
\label{eq:methods_doubled_string_action}
\end{equation}
For Hermitian Pauli strings with $\mathcal{O}_{i,k}^2=I$, the normalized doubled-string expectation is
\begin{align}
\langle \mathcal{O}^{(2)}_{i,k}\rangle_{\rho_{\rm dec}}
&=
\frac{
\langle\langle \rho_{\rm dec}|
\mathcal{O}^{(2)}_{i,k}
|\rho_{\rm dec}\rangle\rangle
}{
\langle\langle \rho_{\rm dec}|\rho_{\rm dec}\rangle\rangle
}
\nonumber\\
&=
\frac{
\Tr[
\rho_{\rm dec}
\mathcal{O}_{i,k}
\rho_{\rm dec}
\mathcal{O}_{i,k}
]
}{
\Tr[
\rho_{\rm dec}^{2}
]
}
\equiv
C^{(2)}_{i,k}.
\label{eq:methods_renyi_string}
\end{align}
This quantity is nonlinear in $\rho_{\rm dec}$ and is therefore sensitive to density-matrix-level order that may be invisible to ordinary single-copy observables.

Although Eq.~\eqref{eq:methods_renyi_string} is nonlinear in one copy of the density matrix, it is measured as a linear observable on two independently prepared replicas.  Let $\rho_A=\rho_B=\rho_{\rm dec}$  denote identical copies oof $\rho_{\rm dec}$ in replicas $A$ and $B$, and let
\begin{equation}
S_{AB}
=
\prod_{j=1}^{L}
S_{j_Aj_B}
\label{eq:methods_global_swap_operator}
\end{equation}
be the global SWAP operator between replicas $A$ and $B$.  We use the identity
\begin{equation}
 {\Tr(AB)=\sum_{j_A,j_B}\langle j_A,j_B|A\otimes B|j_B,j_A\rangle=\Tr[
(A\otimes B)S_{AB}
]}
\label{eq:methods_swap_trace_identity}
\end{equation}
for arbitrary operators $A$ and $B$ on the same Hilbert space, where the trace on the LHS is performed over a single copy, while the RHS trace is over both copies.

The denominator of Eq.~\eqref{eq:methods_renyi_string} is therefore
\begin{align}
D
&=
\Tr(\rho_{\rm dec}^{2})
\nonumber\\
&=
\Tr[
(\rho_A\otimes\rho_B)S_{AB}
].
\label{eq:methods_swap_denominator}
\end{align}
The numerator of Eq.~\eqref{eq:methods_renyi_string},  {denoted by $N_{i,k}$}, can be obtained in two equivalent ways.  First, using Eq.~\eqref{eq:methods_swap_trace_identity}, we have
\begin{align}
N_{i,k}
&=
\Tr[
\rho_{\rm dec}
\mathcal{O}_{i,k}
\rho_{\rm dec}
\mathcal{O}_{i,k}
]
\nonumber\\
&=
\Tr[
(\rho_A\mathcal{O}_{i,k})
\otimes
(\rho_B\mathcal{O}_{i,k})
S_{AB}
]
\nonumber\\
&=
\Tr[
(\rho_A\otimes\rho_B)
(\mathcal{O}^{A}_{i,k}\otimes\mathcal{O}^{B}_{i,k})
S_{AB}
].
\label{eq:methods_swap_numerator_identity}
\end{align}
{Second, in the circuit implementation, we can equivalently} insert the Pauli strings on only one replica, i.e. $A$, immediately before the SWAP readout.  Since $\mathcal{O}_{i,k}$ is Hermitian
this gives
\begin{align}
\langle S_{AB}\rangle_{\mathcal{O}^{A}{\rm\ inserted}}
&=
\Tr[
(
\mathcal{O}^{A}_{i,k}
\rho_A
\mathcal{O}^{A}_{i,k}
\otimes
\rho_B
)
S_{AB}
]
\nonumber\\
&=
\Tr[
\mathcal{O}_{i,k}
\rho_{\rm dec}
\mathcal{O}_{i,k}
\rho_{\rm dec}
]
\nonumber\\
&=
\Tr[
\rho_{\rm dec}
\mathcal{O}_{i,k}
\rho_{\rm dec}
\mathcal{O}_{i,k}
]
=
N_{i,k}.
\label{eq:methods_one_replica_insertion}
\end{align}
Thus one does not need to insert Pauli strings into both replicas in the actual circuit \CH{implementation of $N_{i,k}$}.  Applying $\mathcal{O}_{i,k}$ to one copy before the destructive SWAP measurement is sufficient.

We next give the details of the destructive SWAP readout used to evaluate the
two-replica diagnostic. For two replicas $A$ and $B$,  we express the global SWAP operator
in terms of local SWAPs,
\begin{equation}
S_{AB}
=
\prod_{j=1}^{L}S_{j_Aj_B}.
\label{eq:methods_global_swap}
\end{equation}
This operator satisfies
\begin{equation}
\Tr\!\left[
(\rho_A\otimes\rho_B)S_{AB}
\right]
=
\Tr(\rho_A\rho_B).
\label{eq:methods_swap_identity}
\end{equation}
Thus, when the two replicas are prepared in the same mixed state,
$\rho_A=\rho_B=\rho_{\rm dec}$, the expectation value of $S_{AB}$ directly
gives the purity,
\begin{equation}
\langle S_{AB}\rangle
=
\Tr(\rho_{\rm dec}^{2}).
\label{eq:methods_swap_purity}
\end{equation}

To measure $S_{AB}$ destructively, each corresponding pair
$(j_A,j_B)$ is rotated from the Bell basis to the computational basis by
\begin{equation}
{\rm CNOT}_{j_A\rightarrow j_B}
\quad {\rm followed\ by}\quad
H_{j_A},
\label{eq:methods_destructive_swap_pair}
\end{equation}
after which both qubits are measured. In our implementation, the CNOT is
compiled using the available CZ interaction as
\begin{equation}
{\rm CNOT}_{j_A\rightarrow j_B}
=
H_{j_B}
{\rm CZ}_{j_A,j_B}
H_{j_B}.
\label{eq:methods_cnot_from_cz}
\end{equation}
The Bell states are eigenstates of the local SWAP operator
$S_{j_Aj_B}$: the three symmetric Bell states have eigenvalue $+1$, whereas
the antisymmetric singlet has eigenvalue $-1$.

Under the Bell-basis rotation
in Eq.~\eqref{eq:methods_destructive_swap_pair}, the singlet is mapped to the
computational-basis outcome
$(b_{j,A},b_{j,B})=(1,1)$, while the other three outcomes correspond to
SWAP eigenvalue $+1$. Therefore, for each measured pair,
\begin{equation}
s_j
=
(-1)^{b_{j,A}b_{j,B}},
\qquad
b_{j,A},b_{j,B}\in\{0,1\},
\label{eq:methods_local_swap_estimator}
\end{equation}
is a single-shot estimator of the local SWAP eigenvalue. Since the global SWAP
is the product of the local SWAPs, the corresponding estimator for one
experimental shot is
\begin{equation}
s_{AB}
=
\prod_{j=1}^{L}s_j
=
(-1)^{
\sum_{j=1}^{L}
b_{j,A}b_{j,B}
}.
\label{eq:methods_global_swap_estimator}
\end{equation}
Averaging $s_{AB}$ over repeated shots therefore gives the expectation value
of $S_{AB}$.

For the denominator of the R\'enyi-2 string correlator, both replicas are
measured directly with this destructive SWAP circuit, giving
\begin{equation}
D
=
\Tr(\rho_{\rm dec}^{2})
=
\langle s_{AB}\rangle_{\rm no\ string}.
\label{eq:methods_swap_denominator_average}
\end{equation}
For the numerator, the string operator $\mathcal{O}_{i,k}$ is first applied
to replica $A$, after which the same destructive SWAP measurement is
performed. Using
$S_{AB}\mathcal{O}_{i,k}^{A}
=
\mathcal{O}_{i,k}^{B}S_{AB}$,
we have
\begin{equation}
N_{i,k}
=
\Tr\!\left[
\rho_{\rm dec}\mathcal{O}_{i,k}
\rho_{\rm dec}\mathcal{O}_{i,k}
\right]
=
\langle s_{AB}\rangle_{\rm string}.
\label{eq:methods_swap_numerator_average}
\end{equation}
The measured two-replica diagnostic is therefore
\begin{equation}
C^{(2)}_{i,k}
=
\frac{N_{i,k}}{D}
=
\frac{
\langle s_{AB}\rangle_{\rm string}
}{
\langle s_{AB}\rangle_{\rm no\ string}
}.
\label{eq:methods_measured_doubled_string_ratio}
\end{equation}

We then show why $C^{(2)}_{i,k}=1$ for the ideal decohered cluster
construction. Using the
trajectory decomposition, we have
\begin{align}
D
&=
\Tr(\rho_{\rm dec}^{2})
\nonumber\\
&=
\sum_{\bm{\eta},\bm{\eta}'}
P(\bm{\eta})P(\bm{\eta}')
\left|
\langle\psi_{\bm{\eta}}|
\psi_{\bm{\eta}'}\rangle
\right|^{2},
\label{eq:methods_denominator_trajectory_general}
\end{align}
whereas
\begin{align}
N_{i,k}
&=
\Tr\!\left[
\rho_{\rm dec}\mathcal{O}_{i,k}
\rho_{\rm dec}\mathcal{O}_{i,k}
\right]
\nonumber\\
&=
\sum_{\bm{\eta},\bm{\eta}'}
P(\bm{\eta})P(\bm{\eta}')
s_{\bm{\eta}}s_{\bm{\eta}'}
\left|
\langle\psi_{\bm{\eta}}|
\psi_{\bm{\eta}'}\rangle
\right|^{2}.
\label{eq:methods_numerator_trajectory_general}
\end{align}

For the cluster graph state, distinct $Z$-dephasing trajectories are
orthogonal. To see this, recall that
$K_j=Z_{j-1}X_jZ_{j+1}$ and
$K_j|\psi_{\rm cl}\rangle=|\psi_{\rm cl}\rangle$. A local $Z_j$ anticommutes
with $K_j$, while commuting with all
$K_{\ell\neq j}$. Hence,
\begin{equation}
K_j Z_j|\psi_{\rm cl}\rangle
=
-Z_j|\psi_{\rm cl}\rangle,
\end{equation}
so applying $Z_j$ flips only the eigenvalue of the corresponding stabilizer
from $+1$ to $-1$. More generally, for a dephasing trajectory
$|\psi_{\bm{\eta}}\rangle=Z(\bm{\eta})|\psi_{\rm cl}\rangle$, we have
\begin{equation}
K_j|\psi_{\bm{\eta}}\rangle
=
(-1)^{\eta_j}|\psi_{\bm{\eta}}\rangle,
\qquad
j\in\mathcal{D}_{\rm even}.
\label{eq:methods_trajectory_stabilizer_sector}
\end{equation}
Thus, each trajectory $\bm{\eta}$ is labeled by a definite pattern of
stabilizer eigenvalues. If $\bm{\eta}\neq\bm{\eta}'$, there is at least one
dephased site $j$ for which $\eta_j\neq\eta_j'$, so the two trajectory states
have opposite eigenvalues of the same Hermitian stabilizer $K_j$. They must
therefore be orthogonal,
\begin{equation}
\langle\psi_{\bm{\eta}}|\psi_{\bm{\eta}'}\rangle
=
0,
\qquad
\bm{\eta}\neq\bm{\eta}'.
\end{equation}

Following these, we get
\begin{align}
D
&=
\sum_{\bm{\eta}}
P(\bm{\eta})^{2},
\label{eq:methods_d_fixed_point}
\\
N_{i,k}
&=
\sum_{\bm{\eta}}
P(\bm{\eta})^{2}s_{\bm{\eta}}^{2}
=
\sum_{\bm{\eta}}
P(\bm{\eta})^{2},
\label{eq:methods_n_fixed_point}
\end{align}
because $s_{\bm{\eta}}=\pm1$ and hence $s_{\bm{\eta}}^{2}=1$. Therefore,
\begin{equation}
C^{(2)}_{i,k}
=
\frac{N_{i,k}}{D}
=
1.
\label{eq:methods_renyi_string_fixed_point}
\end{equation}
The distinction from the conventional single-copy string correlator is thus
transparent. The latter directly averages the signed trajectory values
$s_{\bm{\eta}}$, which can cancel in the mixed-state ensemble, whereas the
two-replica diagnostic depends on $s_{\bm{\eta}}^{2}$ and is therefore
insensitive to this trajectory-sign cancellation.

\section{Half-chain entanglement spectrum under dephasing}
\label{sec:methods_entanglement_spectrum}  

We provide details of the half-chain entanglement-spectrum analysis discussed in the main text.  For a density matrix $\rho$, the reduced density matrix of subsystem $A$ is
\begin{equation}
\rho_A
=
\Tr_{\bar A}\rho ,
\label{eq:methods_half_chain_reduced_density_matrix}
\end{equation}
where $\bar A$ denotes the complement of $A$.  If $\lambda_\alpha$ are the eigenvalues of $\rho_A$, the corresponding entanglement levels are defined as
\begin{equation}
\xi_\alpha
=
-\log\lambda_\alpha .
\label{eq:methods_entanglement_levels}
\end{equation}

We first consider the ideal cluster graph state. Let $\mathcal{S}$ denote its stabilizer group, generated by $L$ independent mutually commuting Pauli operators $K_j$. The subgroup containing only stabilizers fully supported inside $A$ is denoted by $\mathcal{S}_A$, 
\begin{equation} 
\mathcal{S}_A = \langle G_1,\ldots,G_{r_A}\rangle , \label{eq:methods_reduced_stabilizer_group} 
\end{equation} where $G_\ell$ are independent commuting generators and $r_A$ is their number. Generally, any pure stabilizer state specified by $L$ independent commuting Pauli stabilizers can be written as \begin{equation} \rho_{\rm cl} = |\psi_{\rm cl}\rangle\langle\psi_{\rm cl}| = \frac{1}{2^L} \sum_{g\in\mathcal{S}} g . \label{eq:methods_full_stabilizer_density_matrix} \end{equation} For the cluster state, $\mathcal{S}$ is generated by the cluster stabilizers defined above.

Thus, only stabilizers supported entirely within
$A$ contribute to the reduced density matrix. We denote this subgroup by
\begin{equation}
\mathcal{S}_A
=
\left\{
g\in\mathcal{S}\,\middle|\,{\rm supp}(g)\subseteq A
\right\}.
\label{eq:methods_subsystem_stabilizer_group}
\end{equation}
For $g\in\mathcal{S}_A$, one can write
$g=g_A\otimes I_{\bar A}$, such that
$\Tr_{\bar A}(g)=2^{|\bar A|}g_A$. Using
$L=|A|+|\bar A|$, we therefore obtain
\begin{equation}
\rho_A^{\rm cl}
=
\frac{1}{2^{|A|}}
\sum_{g\in\mathcal{S}_A} g_A .
\label{eq:methods_reduced_stabilizer_sum}
\end{equation}
Let $\mathcal{S}_A=\langle G_1,\ldots,G_{r_A}\rangle$, where
$G_\ell$ are $r_A$ independent commuting stabilizer generators. The projector
onto their simultaneous $+1$ eigenspace is
\begin{equation}
P_A^{(0)}
=
\prod_{\ell=1}^{r_A}
\frac{I+G_\ell}{2}.
\label{eq:methods_reduced_stabilizer_projector}
\end{equation}
Expanding this projector gives
\begin{equation}
P_A^{(0)}
=
\frac{1}{2^{r_A}}
\sum_{g\in\mathcal{S}_A} g_A ,
\label{eq:methods_projector_expansion}
\end{equation}
and Eq.~\eqref{eq:methods_reduced_stabilizer_sum} can consequently be written as
\begin{equation}
\rho_A^{\rm cl}
=
\frac{1}{2^{|A|-r_A}}
P_A^{(0)}.
\label{eq:methods_reduced_stabilizer_density_matrix}
\end{equation}

This form has a simple interpretation. The $r_A$ independent commuting
stabilizers fully supported within $A$ impose $r_A$ independent constraints on
the $|A|$ qubits of the subsystem. The simultaneous $+1$ eigenspace therefore
has dimension
\begin{equation}
{\rm rank}\,P_A^{(0)}
=
2^{|A|-r_A}.
\label{eq:methods_projector_rank}
\end{equation}
Since $\rho_A^{\rm cl}$ is proportional to the projector
$P_A^{(0)}$, it is maximally mixed within this subspace. Its
$2^{|A|-r_A}$ nonzero eigenvalues are therefore all equal,
\begin{equation}
\lambda_\alpha
=
2^{-(|A|-r_A)},
\qquad
\alpha=1,\ldots,2^{|A|-r_A},
\label{eq:methods_reduced_stabilizer_eigenvalues}
\end{equation}
with all remaining eigenvalues equal to zero.

{As a warm-up, we first consider} the half-chain cut of the one-dimensional cluster state used here. There is one entanglement cut and hence one unresolved virtual boundary degree of freedom.  In stabilizer terms, this means that one qubit inside $A$ is not fixed by the stabilizers fully supported in $A$:
\begin{equation}
|A|-r_A=1 .
\label{eq:methods_rank_condition}
\end{equation}
Thus
\begin{equation}
\rho_A^{\rm cl}
=
\frac{1}{2}P_A^{(0)},
\label{eq:methods_cluster_half_chain_rank_two}
\end{equation}
where $P_A^{(0)}$ has rank two.  Consequently, the reduced density matrix corresponding to the half-chain cut has two equal nonzero eigenvalues,
\begin{equation}
\lambda_1=\lambda_2=\frac{1}{2},
\label{eq:methods_cluster_half_chain_eigenvalues}
\end{equation}
up to the normalization convention for the chosen subsystem.

{More generally}, we now consider the effect of the engineered even-sublattice dephasing channel on the half-chain entanglement spectrum.  For a fixed dephasing trajectory, the output state is a locally $Z$-dressed cluster state,
\begin{equation}
\ket{\psi_{\bm{\eta}}}
=
Z(\bm{\eta})\ket{\psi_{\rm cl}},
\qquad
Z(\bm{\eta})
=
\prod_{j\in{\rm even}}
Z_j^{\eta_j},
\qquad
\eta_j\in\{0,1\}.
\label{eq:methods_trajectory_z_dressed_cluster}
\end{equation}
With respect to the bipartition $A\cup\bar A$, the trajectory operator factorizes as
\begin{equation}
Z(\bm{\eta})
=
Z_A(\bm{\eta}_A)
\otimes
Z_{\bar A}(\bm{\eta}_{\bar A}) .
\label{eq:methods_trajectory_z_factorization}
\end{equation}
Therefore, after tracing out $\bar A$, the components acting only on $\bar A$ drop out by unitary invariance of the partial trace.  The reduced density matrix for a single trajectory is
\begin{equation}
\rho_A(\bm{\eta})
=
Z_A(\bm{\eta}_A)
\rho_A^{\rm cl}
Z_A(\bm{\eta}_A).
\label{eq:methods_trajectory_reduced_density_matrix}
\end{equation}
Thus each dephasing trajectory is related to the fixed-point reduced density matrix by a unitary conjugation acting entirely within subsystem $A$.  This conjugation can rotate the eigenvectors of $\rho_A^{\rm cl}$, but it cannot change its eigenvalues.  Consequently, every individual trajectory has the same nonzero half-chain entanglement spectrum as the original cluster state.

The remaining question is whether the incoherent average over trajectories can split the twofold edge degeneracy. Given $\rho_A^{\rm cl}
=
\frac{1}{2}P_A^{(0)},
P_A^{(0)}
=
\prod_{\ell=1}^{r_A}
\frac{1+G_\ell}{2}$, and the Pauli operator $Z_A(\bm{\eta}_A)$, which either commutes or anticommutes with each reduced stabilizer generator $G_\ell$, we define the corresponding syndrome signs by
\begin{equation}
Z_A(\bm{\eta}_A)
G_\ell
Z_A(\bm{\eta}_A)
=
\sigma_\ell(\bm{\eta}_A)G_\ell,
\qquad
\sigma_\ell(\bm{\eta}_A)=\pm1,
\label{eq:methods_stabilizer_syndrome_sign}
\end{equation}
The corresponding syndrome vector is
\begin{equation}
\bm{\sigma}(\bm{\eta}_A)
=
\{
\sigma_1(\bm{\eta}_A),
\ldots,
\sigma_{r_A}(\bm{\eta}_A)
\}
\label{eq:methods_syndrome_vector}
\end{equation}
which labels the reduced stabilizer syndrome produced by the trajectory.  We then obtain
\begin{align}
\rho_A(\bm{\eta})
&=
Z_A(\bm{\eta}_A)
\left[
\frac{1}{2}P_A^{(0)}
\right]
Z_A(\bm{\eta}_A)
\nonumber\\
&=
\frac{1}{2}
\prod_{\ell=1}^{r_A}
\frac{
1+\sigma_\ell(\bm{\eta}_A)G_\ell
}{2}.
\label{eq:methods_trajectory_syndrome_derivation}
\end{align}
Defining the syndrome projector
\begin{equation}
P_A^{(\bm{\sigma})}
=
\prod_{\ell=1}^{r_A}
\frac{1+\sigma_\ell G_\ell}{2},
\label{eq:methods_syndrome_projector}
\end{equation}
the trajectory-reduced density matrix becomes
\begin{equation}
\rho_A(\bm{\eta})
=
\frac{1}{2}
P_A^{(\bm{\sigma}(\bm{\eta}_A))}.
\label{eq:methods_trajectory_syndrome_density_matrix}
\end{equation}

This equation is the key point.  A dephasing trajectory changes the reduced stabilizer syndrome sector, but it does not reduce the rank of the sector.  Each projector $P_A^{(\bm{\sigma})}$ has the same rank as $P_A^{(0)}$, namely rank two for the half-chain cut.  Thus the trajectory changes the bulk stabilizer label while leaving the two-dimensional virtual-edge subspace unresolved.

After the trajectory record is discarded, the reduced density matrix is a classical mixture over these reduced stabilizer-syndrome sectors:
\begin{equation}
\rho_A^{\rm dec}
=
\sum_{\bm{\sigma}}
W_{\bm{\sigma}}
\frac{P_A^{(\bm{\sigma})}}{2}.
\label{eq:methods_dephased_syndrome_mixture}
\end{equation}
Here
\begin{equation}
W_{\bm{\sigma}}
=
\sum_{\bm{\eta}_A:\,\bm{\sigma}(\bm{\eta}_A)=\bm{\sigma}}
P_A(\bm{\eta}_A)
\label{eq:methods_syndrome_weight}
\end{equation}
is the total probability of all dephasing trajectories that produce the same reduced syndrome $\bm{\sigma}$ inside subsystem $A$.

 Since each block $P_A^{(\bm{\sigma})}/2$ has rank two, the nonzero eigenvalues are
\begin{equation}
\lambda_{\bm{\sigma},1}
=
\lambda_{\bm{\sigma},2}
=
\frac{W_{\bm{\sigma}}}{2}
\label{eq:methods_dephased_entanglement_eigenvalues}
\end{equation}
for every populated syndrome sector $\bm{\sigma}$.

For the $L=8$ half-chain bipartition used here, we take
$A=\{1,2,3,4\}$. The reduced cluster-state density matrix is
\begin{equation}
\rho_A^{\rm cl}
=
\frac{1}{16}
(I+K_1)(I+K_2)(I+K_3).
\label{eq:methods_l8_cluster_reduced_density}
\end{equation}
Although the dephasing channel acts on the two even sites $j=2$ and $j=4$
within $A$, only $Z_2$ changes a stabilizer syndrome visible in the reduced
density matrix. Specifically, $Z_2$ anticommutes with $K_2$, whereas $Z_4$
commutes with all stabilizers fully supported within $A$. The latter therefore
leaves $\rho_A^{\rm cl}$ invariant. Dephasing operations acting entirely in
$\bar A$ likewise disappear under the partial trace. The reduced density
matrix after dephasing consequently becomes
\begin{equation}
\rho_A^{\rm dec}
=
\frac{1}{16}
(I+K_1)
\left[I+(1-2p)K_2\right]
(I+K_3).
\label{eq:methods_l8_dephased_reduced_density}
\end{equation}
Equivalently, it consists of two reduced-stabilizer sectors with weights
$1-p$ and $p$. Since each sector has rank two, the four nonzero eigenvalues
are
\begin{equation}
\lambda_1=\lambda_2=\frac{1-p}{2},
\qquad
\lambda_3=\lambda_4=\frac{p}{2}.
\label{eq:methods_l8_dephased_eigenvalues}
\end{equation}
The corresponding entanglement levels are therefore
\begin{equation}
\xi_{1,2}
=
-\log\!\left(\frac{1-p}{2}\right),
\qquad
\xi_{3,4}
=
-\log\!\left(\frac{p}{2}\right).
\label{eq:methods_l8_dephased_levels}
\end{equation}
For the dephasing strength $p=0.1$ used in the main-text comparison, this gives
\begin{equation}
\lambda_{1,2}=0.45,
\qquad
\lambda_{3,4}=0.05,
\end{equation}
or, equivalently [see Fig.~7 in the main text],
\begin{equation}
\xi_{1,2}
=
-\ln(0.45)
\simeq 0.799,
\qquad
\xi_{3,4}
=
-\ln(0.05)
\simeq 2.996.
\label{eq:methods_l8_p01_entanglement_levels}
\end{equation}
Thus, the engineered dephasing generates an additional reduced-stabilizer
sector and shifts the corresponding entanglement levels, while preserving the
twofold pairing within each populated sector.

\section{Measurement-and-feedback realization of mixed-state SPT order}
\label{sec:methods_feedback_mspt}

\begin{figure}
    \centering
    \includegraphics[width=0.7\linewidth]{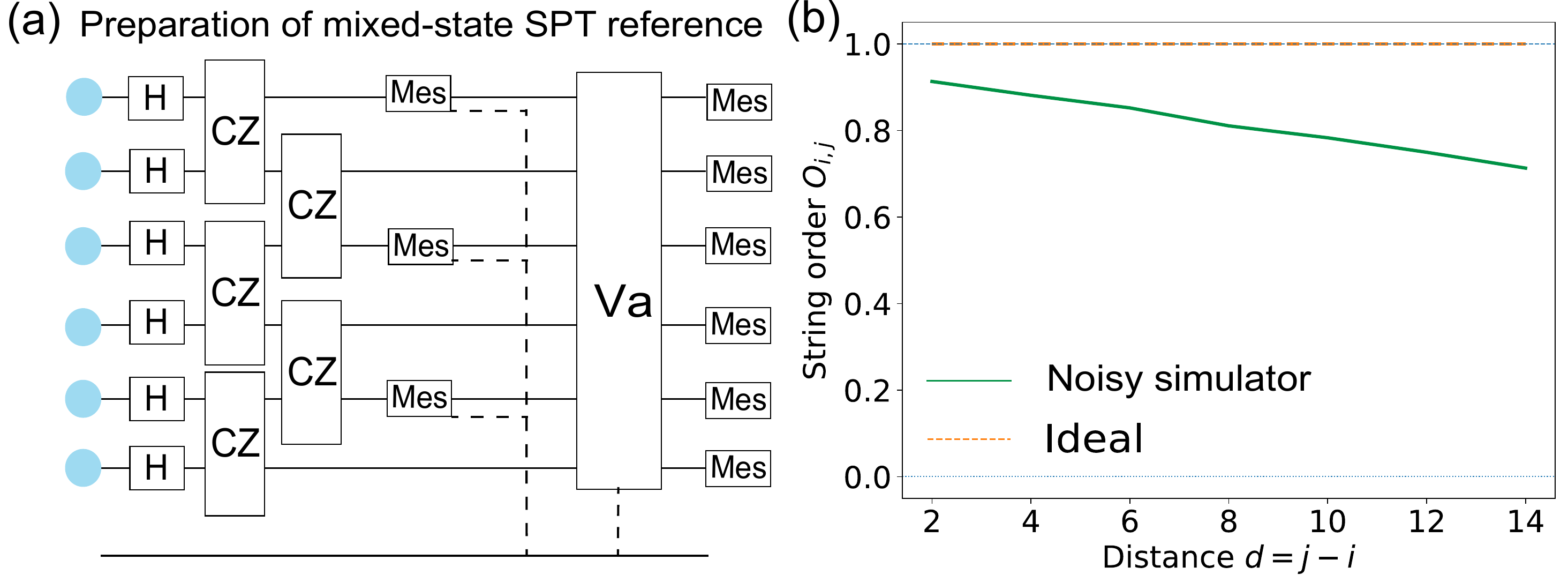}
    \caption{
    \textbf{Measurement-and-feedback realization of mixed-state SPT order.}
    \textbf{a}, Circuit for preparing the feedback-assisted mixed-state SPT.
    Starting from the cluster state, the even-sublattice qubits are measured in
    the $X$ basis, and the resulting measurement record determines the
    conditional feedback unitary $V_\alpha$ in
    Eq.~\eqref{eq:feedback_unitary}.
    \textbf{b}, Long-range order obtained from the protocol in \textbf{a}.
    The plotted quantity is the feedback-mapped string correlator, equivalently
    the odd-sublattice two-point correlator
    $\langle Z_{2i+1}Z_{2j+1}\rangle$ in
    Eq.~\eqref{eq:string_to_zz_mapping}, shown as a function of the endpoint
    separation $d=j-i$ for an $L=16$ chain using a local noisy simulator.
    }
    \label{fig:mspt}
\end{figure}

It is important to distinguish the decoherence-induced ASPT state studied in
the main text from more general mixed-state SPT states, which need not exhibit
the same strong--average symmetry structure. As a complementary reference, we
consider the measurement-and-feedback construction of
Ref.~\cite{luMixedStateLongRange2023}, shown in Fig.~\ref{fig:mspt}
\cite{piroliQuantumCircuitsAssisted2021,
luMeasurementShortcutLongRange2022,tantivasadakarnLongRangeEntanglement2024}.
This protocol starts from the same one-dimensional cluster SPT state but treats
the measurement record differently from the ASPT protocol. In the latter, the
trajectory record associated with the engineered dephasing is averaged over,
whereas here the measurement outcomes are actively used to apply a conditional
feedback operation. The purpose of this comparison is therefore to illustrate
a distinct route to mixed-state SPT order in which hidden cluster string order
is converted into an ordinary long-range correlation.

Concretely, starting from
$\rho_{\rm cl}=|\psi_{\rm cl}\rangle\langle\psi_{\rm cl}|$, every
even-sublattice qubit is measured in the $X$ basis. We denote the measurement
record by
$\alpha=\{\alpha_i\}$, where $\alpha_i=\pm1$ is the outcome of measuring
$X_{2i}$. The projector associated with a given record is
\begin{equation}
P_\alpha
=
\prod_i
\frac{1+\alpha_i X_{2i}}{2}.
\label{eq:measurement_projector}
\end{equation}
Without feedback, different measurement records generate
trajectory-dependent signs in the remaining cluster correlations. To remove
these signs, the measurement record is processed classically and used to
determine the conditional unitary
\begin{equation}
V_\alpha
=
\prod_i
X_{2i+1}^{\frac{1-\prod_{m\le i}\alpha_m}{2}}.
\label{eq:feedback_unitary}
\end{equation}
Thus, an $X$ operation is applied to the odd site $2i+1$ whenever the
cumulative measurement parity
$\prod_{m\le i}\alpha_m$ is $-1$. The feedback therefore compensates the
measurement-dependent sign accumulated along the corresponding cluster string.

For a fixed measurement record $\alpha$, the conditional system state is
\begin{equation}
\rho_\alpha^{\rm fb}
=
V_\alpha
P_\alpha\rho_{\rm cl}P_\alpha
V_\alpha^\dagger.
\label{eq:feedback_conditional_state}
\end{equation}
Averaging over all measurement outcomes gives the feedback-prepared mixed
state
\begin{equation}
\rho_{\rm fb}
=
\sum_\alpha
V_\alpha P_\alpha
\rho_{\rm cl}
P_\alpha V_\alpha^\dagger .
\label{eq:feedback_density_matrix}
\end{equation}
Here the probabilities of the measurement outcomes are already contained in
the unnormalized projected states
$P_\alpha\rho_{\rm cl}P_\alpha$, so no additional probability factor is
required in Eq.~\eqref{eq:feedback_density_matrix}.

The effect of the feedback is particularly transparent in the cluster string
order. For the nonlocal string operator $\mathcal{O}_{i,j}$ defined in
Eq.~\eqref{eq:main_cluster_string}, the measurement outcomes of the intervening
even-sublattice $X$ operators determine the trajectory-dependent sign of the
string. The feedback unitary $V_\alpha$ removes this sign and maps the hidden
string order onto an ordinary two-point correlation between the remaining odd
sites. Consequently,
\begin{equation}
\begin{aligned}
C^{\rm fb}_{i,j}
&=
\Tr\!\left[
\rho_{\rm fb}
Z_{2i+1}Z_{2j+1}
\right]
\\
&=
\Tr\!\left[
\rho_{\rm cl}\,
\mathcal{O}_{i,j}
\right].
\end{aligned}
\label{eq:string_to_zz_mapping}
\end{equation}
The nonlocal string order of the input cluster SPT can therefore be detected
after feedback through a conventional single-copy long-range $ZZ$ correlator
\cite{luMixedStateLongRange2023,piroliQuantumCircuitsAssisted2021,
luMeasurementShortcutLongRange2022,tantivasadakarnLongRangeEntanglement2024}.

Figure~\ref{fig:mspt}(b) shows the resulting correlator for an $L=16$ chain.
The feedback-mapped $ZZ$ correlation remains finite as the endpoint separation increases, demonstrating that the measurement record can be used to recover
long-range order from the underlying cluster string structure even after
averaging over measurement outcomes. This provides a useful comparison with
the decoherence-induced ASPT protocol. In the feedback construction, the
classical record is retained and actively used to remove the
trajectory-dependent signs, allowing the mixed-state order to be detected by a
single-copy correlator.

\end{document}